\documentclass[aps]{revtex4}
\usepackage{amsmath}
\usepackage{graphicx}
\usepackage{dcolumn}
\usepackage{bm}

\begin{document}

\title{Mini-review: Spatial solitons supported by localized gain}
\author{Boris A. Malomed}
\address{Department of Physical Electronics, School of Electrical
Engineering, Faculty of Engineering, Tel Aviv University, Tel Aviv
69978, Israel}

\begin{abstract}
The creation of stable 1D and 2D localized modes in lossy nonlinear
media is a fundamental problem in optics and plasmonics. This
article gives a short review of theoretical methods elaborated for
this purpose, using localized gain applied at one or several ``hot
spots" (HSs). The introduction surveys a broad class of models for
which this approach was developed. Other sections focus in some
detail on basic 1D continuous and discrete systems, where the
results can be obtained, partly or fully, in an analytical form
(verified by comparison with numerical results), which provides a
deeper insight into the nonlinear dynamics of optical beams in
dissipative nonlinear media. In particular, considered are the
single and double HS in the usual waveguide with the self-focusing
(SF) or self-defocusing (SDF) Kerr nonlinearity, which gives rise to
rather sophisticated results, in
spite of apparent simplicity of the model; solitons attached to a $\mathcal{%
PT}$-symmetric dipole embedded into the SF or SDF medium; gap solitons
pinned to an HS in a Bragg grating (BG); and discrete solitons in a 1D
lattice with a ``hot site". \smallskip \noindent

\textit{List of acronyms}: 1D -- one-dimensional; 2D -- two-dimensional;
b.c. -- boundary conditions; BG -- Bragg grating; CGLE -- complex
Ginzburg-Landau equation; CME -- coupled-mode equations; HS -- hot spot;
NLSE -- nonlinear Schr\"{o}dinger equation; $\mathcal{PT}$ -- parity-time
(symmetry); SDS -- spatial dissipative soliton; SF -- self-focusing; SDF --
self-defocusing; VK -- Vakhitov-Kolokolov (stability criterion); WS -- warm
spot

\smallskip \noindent OCIS numbers: 190.6135; 240.6680; 230.4480; 190.4360
\end{abstract}

\maketitle

\section{Introduction and basic models}

Spatial dissipative solitons (SDSs) are self-trapped beams of light \cite%
{Rosanov,MT,Akhmed} or plasmonic waves \cite{plas1}-\cite{Marini}
propagating in planar or bulk waveguides. They result from the balance
between diffraction and self-focusing (SF) nonlinearity, which is maintained
simultaneously with the balance between the material loss and compensating
gain. Due to their basic nature, SDSs are modes of profound significance to
nonlinear photonics (optics and plasmonics), as concerns the fundamental
studies and potential applications alike. In particular, a straightforward
possibility is to use each sufficiently narrow SDS beams as signal carriers
in all-optical data-processing schemes. This application, as well as other
settings in which the solitons occur, stresses the importance of the
stabilization of the SDSs modes, and of development of enabling techniques
for the generation and steering of such planar and bulk \ beams.

In terms of the theoretical description, basic models of the SDS dynamics
make use of complex Ginzburg-Landau equations (CGLEs). The prototypical one
is the CGLE with the cubic nonlinearity, which includes the conservative
paraxial-diffraction and Kerr terms, nonlinear (cubic) loss with coefficient
$\epsilon >0$, which represents two-photon absorption in the medium, and the
spatially uniform linear gain, with strength $\gamma >0$, aiming to
compensate the loss \cite{Rosanov,MT}:

\begin{equation}
\frac{\partial u}{\partial z}=\frac{i}{2}\nabla _{\perp }^{2}u-\left(
\epsilon -i\beta \right) |u|^{2}u+\gamma u.  \label{GL}
\end{equation}%
Here $u$ is the complex amplitude of the electromagnetic wave in the spatial
domain, $z$ is the propagation distance, the paraxial-diffraction operator $%
\nabla _{\perp }^{2}$ acts on transverse coordinates $\left( x,y\right) $ in
the case of the propagation in the bulk, or on the single coordinate, $x$,
in the planar waveguide. Accordingly, Eq. (\ref{GL}) is considered as two-
or one-dimensional (2D or 1D) equation in those two cases. The equation is
normalized so that the diffraction coefficient is $1$, while $\beta $ is the
Kerr coefficient, $\beta >0$ and $\beta <0$ corresponding to the SF and
self-defocusing (SDF) signs of the nonlinearity, respectively.

A more general version of the CGLE may include an imaginary part of the
diffraction coefficient \cite{CH,CGL,encyclopedia}, which is essential, in
particular, for the use of the CGLE as a model of the traveling-wave
convection \cite{Kolodner,Cross}. However, in optical models that
coefficient, which would represent diffusivity of photons, is usually absent.

A well-known fact is that the 1D version of Eq. (\ref{GL}) gives rise to an
exact solution in the form of an exact \textit{chirped} SDS, which is often
called a \textit{Pereira-Stenflo soliton} \cite{SH,Lennart}:%
\begin{eqnarray}
u\left( x,z\right) =Ae^{ikz}\left[ \mathrm{sech}\left( \kappa x\right) %
\right] ^{1+i\mu },  \label{PS} \\
A^{2}=3\gamma /\left( 2\epsilon \right) ,~\kappa ^{2}=\gamma /\mu ,~k=\left(
\gamma /2\right) \left( \mu ^{-1}-\mu \right) ,  \label{PSconstants}
\end{eqnarray}%
where the chirp coefficient is%
\begin{equation}
\mu =\sqrt{\left( 3\beta /2\epsilon \right) ^{2}+2}-3\beta /\left( 2\epsilon
\right) .  \label{mu}
\end{equation}%
This exact solution is subject to an obvious instability, due to the action
of the uniform linear gain on the zero background far from the soliton's
core. Therefore, an important problem is the design of physically relevant
models which may produce stable SDS.

One possibility is to achieve full stabilization of the solitons in systems
of linearly coupled CGLEs~modeling dual-core waveguides, with the linear
gain and loss acting in different cores~\cite{Dual1}-\cite{Dual6}, \cite%
{Marini}. This includes, \textit{inter alia}, a $\mathcal{PT}$-symmetric
version of the system that features the exact balance between the gain and
loss~\cite{PT1,PT2}. The simplest example of such a stabilization mechanism
is offered by the following coupled CGLE system \cite{Javid}:

\begin{eqnarray}
\frac{\partial u}{\partial z} &=&\frac{i}{2}\nabla _{\perp }^{2}u-\left(
\epsilon -i\beta \right) |u|^{2}u+\gamma u+i\lambda v,  \label{u-c} \\
\frac{\partial v}{\partial z} &=&\left( iq-\Gamma \right) v+i\lambda u,
\label{v-c}
\end{eqnarray}%
where $\lambda $ is the linear-coupling coefficient, $v\left( x,z\right) $
and $\Gamma >0$ are the electromagnetic-wave amplitude and the linear-loss
rate in the stabilizing dissipative core, and $q$ is a possible wavenumber
mismatch between the cores. In the case of $q=0$, the zero background is
stable in the framework of Eqs. (\ref{u-c}) and (\ref{v-c}) under condition%
\begin{equation}
\gamma <\Gamma <\lambda ^{2}/\gamma .  \label{gG}
\end{equation}

The same ansatz (\ref{PS}) which produced the Pereira-Stenflo soliton for
the uncoupled CGLE yields an exact solution of the coupled system (\ref{u-c}%
), (\ref{v-c}):%
\begin{equation}
\left\{ u\left( x,z\right) ,v\left( x,z\right) \right\} =\left\{ A,B\right\}
e^{ikz}\left[ \mathrm{sech}\left( \kappa x\right) \right] ^{1+i\mu },
\end{equation}%
with chirp $\mu $ given by the same expression (\ref{mu}) as above, and
\begin{equation}
B=i\lambda \left[ \Gamma +i\left( k-q\right) \right] ^{-1}A.  \label{B}
\end{equation}%
A stable soliton is obtained if a pair of distinct solutions are found,
compatible with the condition of the stability for the zero background
[which is Eq. (\ref{gG}) in the case of $q=0$], instead of the single
solution in the case of Eq. (\ref{PSconstants}). Then, the soliton with the
larger amplitude is stable, coexisting, as an \textit{attractor}, with the
stable zero solution, while the additional soliton with a smaller amplitude
plays the role of an unstable \textit{separatrix} which delineates the
boundary between attraction basins of the two coexisting stable solutions
\cite{Javid}.

In the case of $q=0$, the aforementioned condition of the existence of two
solutions reduces to
\begin{equation}
\gamma \Gamma \left( 1-\mu ^{2}\right) >4\mu ^{2}\left[ \left( \lambda
^{2}-\gamma \Gamma \right) +2\Gamma \left( \Gamma -\gamma \right) \right] .
\label{two}
\end{equation}%
In particular, it follows from Eq. (\ref{two}) and (\ref{mu}) that a related
necessary condition, $\mu <1$, implies $\epsilon <3\beta $, i.e., the Kerr
coefficient, $\beta $, must feature the SF sign, and the cubic-loss
coefficient, $\epsilon $, must be sufficiently small in comparison with $%
\beta $. If inequality $\mu <1$ holds, and zero background is close to its
stability boundaries, i.e., $0<\Gamma -\gamma \ll \gamma $ and $0<\lambda
^{2}-\gamma \Gamma \ll \gamma ^{2}$, see Eq. (\ref{gG}), parameters of the
stable soliton with the larger amplitude are%
\begin{equation}
\kappa _{\mathrm{st}}\approx \sqrt{\frac{\gamma }{\mu }}\frac{1-\mu ^{2}}{%
1+\mu ^{2}},A_{\mathrm{st}}\approx \sqrt{\frac{3\gamma }{2\epsilon }}\frac{%
1-\mu ^{2}}{1+\mu ^{2}},B_{\mathrm{st}}\approx \frac{2\mu }{\left( 1-i\mu
\right) ^{2}}A,
\end{equation}%
while propagation constant $k$ is given, in the first approximation, by Eq. (%
\ref{PSconstants}). In the same case, the unstable\textit{\ }separatrix
soliton has a small amplitude and large width, while its chirp keeps the
above value (\ref{mu}):%
\begin{gather}
\kappa _{\mathrm{sep}}^{2}\approx \frac{2\left( \Gamma -\gamma \right) }{%
\gamma \left( 1-\mu ^{2}\right) }\sqrt{\lambda ^{2}-\gamma \Gamma },~k_{%
\mathrm{sep}}\approx \sqrt{\lambda ^{2}-\gamma \Gamma }, \\
A_{\mathrm{sep}}^{2}\approx \frac{3\mu }{2\epsilon }\kappa _{\mathrm{sep}%
}^{2},~B_{\mathrm{sep}}\approx iA_{\mathrm{sep}},
\end{gather}

Getting back to models based on the single CGLE, stable solitons can also be
generated by the equation with cubic gain ``sandwiched" between linear and
quintic loss terms, which corresponds to the following generalization of Eq.
(\ref{GL}):

\begin{equation}
\frac{\partial u}{\partial z}=\frac{i}{2}\nabla _{\perp }^{2}u+\left(
\epsilon _{3}+i\beta _{3}\right) |u|^{2}u-\left( \epsilon _{5}+i\beta
_{5}\right) |u|^{2}u-\Gamma u,  \label{CQGL}
\end{equation}%
with $\epsilon _{3}>0$, $\epsilon _{5}>0$, $\Gamma >0$, and $\beta _{5}\geq
0 $. The linear loss, represented by coefficient $\Gamma $, provides for the
stability of the zero solution to Eq. (\ref{CQGL}). The cubic-quintic (CQ)
CGLE was first proposed, in a phenomenological form, by Petviashvili and
Sergeev \cite{PS}. Later, it was demonstrated that the CQ model may be
realized in optics as a combination of linear amplification and saturable
absorption~\cite{cqgle2}-\cite{cqgle5}. Stable dissipative solitons
supported by this model were investigated in detail by means of numerical
and analytical methods \cite{CQ1}-\cite{CQ6}.

The subject of the present mini-review is the development of another method
for creating stable localized modes, which makes use of linear gain applied
at a ``hot spot" (HS), i.e. a localized amplifying region embedded into a
bulk lossy waveguide. The experimental technique which allows one to create
localized gain by means of strongly inhomogeneous distributions of dopants
implanted into the lossy waveguide, which produce the gain if pumped by an
external source of light, is well known \cite{Kip}. Another possibility is
even more feasible and versatile: the dopant density may be uniform, while
the external pump beam is focused on the location where the HS should be
created.

Supporting dissipative solitons by the localized gain was first proposed not
in the framework of CGLEs, but for a gap soliton pinned to an HS in a lossy
Bragg grating (BG) \cite{Mak}. In terms of the spatial-domain dynamics, the
respective model is based on the system of coupled-mode equations (CMEs) for
counterpropagating waves, $u\left( x,z\right) $ and $u\left( x,z\right) $,
coupled by the Bragg reflection:%
\begin{eqnarray}
iu_{z}+iu_{x}+v+\left( |u|^{2}+2|v|^{2}\right) u=  \notag \\
-i\gamma u+i\left( \Gamma _{1}+i\Gamma _{2}\right) \delta (x)u,  \label{o1}
\\
iv_{z}-iv_{x}+u+\left( |v|^{2}+2|u|^{2}\right) v=  \notag \\
-i\gamma v+i\left( \Gamma _{1}+i\Gamma _{2}\right) \delta (x)v,  \label{o2}
\end{eqnarray}%
where the tilt of the light beam and the reflection coefficients are
normalized to be $1$, the nonlinear terms account for the self- and
cross-phase modulation induced by the Kerr effect, $\gamma >0$ is the
linear-loss parameter, $\Gamma _{1}>0$ represents the local gain applied at
the HS [$x=0$, $\delta (x)$ being the Dirac's delta-function], and the
imaginary part of the gain coefficient, $\Gamma _{2}\geq 0$, accounts for a
possible attractive potential induced by the HS (it approximates a local
increase of the refractive index around the HS).

As was mentioned in \cite{Mak} too, and for the first time investigated in
detail in \cite{HSexact}, the HS embedded into the usual planar waveguide is
described by the following modification of Eq. (\ref{GL}):

\begin{equation}
\frac{\partial u}{\partial z}=\frac{i}{2}\nabla _{\perp }^{2}u-\left(
\epsilon -i\beta \right) |u|^{2}u-\gamma u+\left( \Gamma _{1}+i\Gamma
_{2}\right) \delta (x)u,  \label{CGHS}
\end{equation}%
where, as well as in Eqs. (\ref{o1}) and (\ref{o2}), $\Gamma _{1}>0$ is
assumed, and the negative sign in front of $\gamma \geq 0$ represents the
linear loss in the bulk waveguide. Another HS model, based on the 1D CGLE
with the CQ nonlinearity, was introduced in \cite{Valery}:%
\begin{equation}
\frac{\partial u}{\partial z}=\frac{i}{2}\frac{\partial ^{2}u}{\partial x^{2}%
}+i|u|^{2}u-i\beta _{5}|u|^{2}u-\gamma u+\Gamma e^{-x^{2}/w^{2}}|u|^{2}u,
\label{CQHS}
\end{equation}%
where $\beta _{5}>0$ represents the quintic self-defocusing term, $\gamma >0$
and $\Gamma >0$ are, as above, strengths of the bulk losses and localized
\emph{cubic} gain, and $w$ is the width of the HS (an approximation
corresponding to $w\rightarrow 0$, with the HS in the form of the
delta-function, may be applied here too). While solitons in uniform media,
supported by the cubic gain, are always unstable against the blowup in the
absence of the quintic loss~\cite{Kramer}, the analysis reported in \cite%
{Valery} demonstrates that, quite counter-intuitively, \emph{stable}
dissipative localized modes in the uniform lossy medium may be supported by
the \emph{unsaturated} localized cubic gain in the model based on Eq. (\ref%
{CQHS}).

In addition to the ``direct" linear gain assumed in the above-mentioned
models, losses in photonic media may be compensated by parametric
amplification, which, unlike the direct gain, is sensitive to the phase of
the signal \cite{param1,param2}. This mechanism can be used for the creation
of a HS, if the parametric gain is applied in a narrow segment of the
waveguide. As proposed in \cite{Ye}, the respective 1D model is based on the
following equation, cf. Eqs. (\ref{CGHS}) and (\ref{CQHS}):%
\begin{equation}
\frac{\partial u}{\partial z}=\frac{i}{2}\frac{\partial ^{2}u}{\partial x^{2}%
}+i|u|^{2}u-\left( \epsilon -i\beta \right) |u|^{2}u-\left( \gamma
-iq\right) u+\Gamma e^{-x^{2}/w^{2}}u^{\ast },  \label{HSparam}
\end{equation}%
where $u^{\ast }$ is the complex conjugate field, $q$ is a real
phase-mismatch parameter, and, as well as in Eq. (\ref{CQHS}), the HS may be
approximated by the delta-function in the limit of $w\rightarrow 0$.

Models combining the localized gain and the uniformly distributed Kerr
nonlinearity and linear loss have been recently developed in various
directions. In particular, 1D models with two or multiple HSs~\cite%
{spotsExact1}-\cite{spotsExact2} and periodic amplifying structures \cite%
{KKV,Yingji}, as well as extended patterns~\cite{Zezyu1,Zezyu2}, have been
studied, chiefly by means of numerical methods. The numerical analysis has
made it also possible to study 2D settings, in which, most notably, stable
localized vortices are supported by the gain confined to an annular-shaped
area~\cite{2D1}-\cite{2D6}. The parametric amplification applied at a ring
may support stable vortices too, provided that the pump 2D beam itself has
an inner vortical structure \cite{Ye2}.

Another ramification of the topic is the development of symmetric
combinations of ``hot" and ``cold" spots, which offer a realization of the
concept of $\mathcal{PT}$-symmetric systems in optical media, that were
proposed and built as settings integrating the balanced spatially separated
gain and loss with a spatially symmetric profile of the local refractive
index \cite{linPT1}-\cite{linPT4}. The study of solitons in nonlinear $%
\mathcal{PT}$-symmetric settings has drawn a great deal of attention \cite%
{nonlinPT1}-\cite{nonlinPT5}, \cite{PT1,PT2}. In particular, it is possible
to consider the 1D model in the form of a symmetric pair of hot and cold
spots described by two delta-functions embedded into a bulk conservative
medium with the cubic nonlinearity \cite{Stuttgart}. A limit case of this
setting, which admits exact analytical solutions for $\mathcal{PT}$%
-symmetric solitons, corresponds to a $\mathcal{PT}$ \textit{dipole}, which
is represented by the derivative of the delta-function in the following 1D
equation \cite{Thawatchai}:%
\begin{equation}
i\frac{\partial u}{\partial z}=-\frac{1}{2}\frac{\partial ^{2}u}{\partial
x^{2}}-\sigma |u|^{2}u-\varepsilon _{0}u\delta \left( x\right) +i\gamma
u\delta ^{\prime }\left( x\right) .  \label{eq}
\end{equation}%
Here $\sigma =+1$ and $-1$ correspond to the SF and SDF bulk nonlinearity,
respectively, $\varepsilon _{0}\geq 0$ is the strength of the attractive
potential, which is a natural conservative component of the $\mathcal{PT}$
\textit{dipole}, and $\gamma $ is the strength of the $\mathcal{PT}$ dipole.

It is also natural to consider discrete photonic settings (lattices), which
appear, in the form of discrete CGLEs, as models of arrayed optical \cite%
{discrCGL1}-\cite{discrCGL10} or plasmonics \cite%
{plasmon-array2,plasmon-array3} waveguides. In this context, lattice
counterparts of the HSs amount to a single \cite{discr} or several \cite%
{discr2} amplified site(s) embedded into a 1D or 2D \cite{discr3} lossy
array. Being interested in tightly localized discrete states, one can
additionally simplify the model by assuming that the nonlinearity is carried
only by the active cores, which gives rise to the following version of the
discrete CGLE, written here in the general 2D form \cite{discr3}:
\begin{eqnarray}
\frac{du_{m,n}}{dz}=\frac{i}{2}\left(
u_{m-1,n}+u_{m+1,n}+u_{m,n-1}+u_{m,n+1}-4u_{m,n}\right)  \notag \\
-\gamma u_{m,n}+\left[ \left( \Gamma _{1}+i\Gamma _{2}\right) +\left(
iB-E\right) |u_{m,n}|^{2}\right] \delta _{m,0}\delta _{n,0}u_{m,n}\;,
\label{eq:gl}
\end{eqnarray}%
where $m,n=0$, $\pm 1$, $\pm 2$, ... are discrete coordinates on the
lattice, $\delta _{m,0}$ and $\delta _{n,0}$ are the Kronecker's symbols,
and the coefficient of the linear coupling between adjacent cores is scaled
to be $1$. As above, $\gamma >0$ is the linear loss in the bulk lattice, $%
\Gamma _{1}>0$ and $\Gamma _{2}\geq 0$ represent the linear gain and linear
potential applied at the HS site ($m=n=0$), while $B$ and $E$ account for
the SF ($B>0$) or SDF ($B<0$) Kerr nonlinearity and nonlinear loss ($E>0$)
or gain ($E<0$) acting at the HS (the unsaturated cubic gain may be a
meaningful feature in this setting \cite{discr3}).

The $\mathcal{PT}$ symmetry can be introduced too in the framework of the
lattice system. In particular, a discrete counterpart of the 1D continuous
model (\ref{eq}) with the $\mathcal{PT}$-symmetric dipole was recently
elaborated in \cite{Raymond}:
\begin{equation}
i{\frac{du_{n}}{dz}}=-\left( C_{n,n-1}u_{n-1}+C_{n+1,n}u_{n+1}\right)
-g_{n}|u_{n}|^{2}u_{n}+i\kappa _{n}u_{n},  \label{DNLSE}
\end{equation}%
where the $\mathcal{PT}$ dimer (discrete dipole) embedded into the
Hamiltonian lattice is represented by $\kappa _{n}=+\kappa $ at $n=0$, $%
-\kappa $ at $n=1$, and $0$ at $n\neq 0,1$. A counterpart of the
delta-functional attractive potential in Eq. (\ref{eq}) corresponds to a
local defect in the inter-site couplings: $C_{1,0}=C_{d}$, $%
C_{n,n-1}=C_{0}\neq C_{d}$ at $n\neq 1$. Lastly, the nonlinearity is assumed
to be carried solely by the dimer embedded into the lattice: $g_{n}=g$ at $%
n=0,1$, and $0$ at $n\neq 0,1$, cf. Eq. (\ref{eq:gl}). Equation (\ref{DNLSE}%
) admits exact analytical solutions for all $\mathcal{PT}$-symmetric and
antisymmetric discrete solitons pinned to the dimer.

In addition to the HS, one can naturally define a ``warm spot" (WS), in the
2D CGLE with the CQ nonlinearity, where the coefficient of the linear loss
is given a spatial profile with a minimum at the WS ($\mathbf{r}=0$) \cite%
{Skarka}. The equation may be taken as the 2D version of Eq. (\ref{CQGL})
with
\begin{equation}
\Gamma (r)=\Gamma _{0}+\Gamma _{2}r^{2},  \label{WS}
\end{equation}%
where $r$ is the radial coordinate, coefficients $\Gamma _{0}$ and $\Gamma
_{2}$ being positive. This seemingly simple model gives rise to a great
variety of stable 2D modes pinned to the WS. Depending on values of
parameters in Eqs. (\ref{CQGL}) and (\ref{WS}), these may be simple
vortices, rotating elliptic, eccentric, and slanted vortices, spinning
crescents, etc. \cite{Skarka}.

Lastly, the use of the spatial modulation of loss coefficients opens another
way for the stabilization of the SDS: as shown in \cite{Barcelona}, the
solitons may be readily made stable if the \emph{spatially uniform} linear
gain is combined with the local strength of the cubic loss, $\epsilon (r)$,
growing from the center to periphery at any rate faster than $r^{D}$, where $%
r$ is the distance from the center and $D$ the spatial dimension. This
setting is described by the following modification of Eq. (\ref{GL}):

\begin{equation}
\frac{\partial u}{\partial z}=\frac{i}{2}\nabla _{\perp }^{2}u-\left[
\epsilon (r)-i\beta \right] |u|^{2}u+\gamma u,  \label{Olga}
\end{equation}%
where, as said above, $\gamma $ and $\epsilon (r)$ are positive, so that $%
\lim_{r\rightarrow \infty }\left( r^{D}/\epsilon (r)\right) =0$, for $D=1$
or $2$.

This mini-review aims to present a survey of basic results obtained for SDSs
pinned to HSs in the class of models outlined above. In view of the limited
length of the article, stress is made on the most fundamental results that
can be obtained in an analytical or semi-analytical form, in combination
with the related numerical findings, thus providing a deep inside into the
dynamics of the underlying photonic systems. In fact, the possibility of
obtaining many essential results in an analytical form is a certain asset of
these models. First, in Section II findings are summarized for the most
fundamental 1D model based on Eq. (\ref{CGHS}), which is followed, in
Section III, by the consideration of the $\mathcal{PT}$-symmetric system (%
\ref{eq}). The BG model (\ref{o1}), (\ref{o2}) is the subject of Section IV,
and Section V deals with the 1D version of the discrete model (\ref{eq:gl}).
The article is concluded by Section VI.

\section{Dissipative solitons pinned to hot spots in the ordinary waveguide}

The presentation in this section is focused on basic model (\ref{CGHS}) and
its extension for two HSs, chiefly following works \cite{HSexact} and \cite%
{spotsExact1}, for the settings with a single and double HS, respectively.
Both analytical and numerical results are presented, which highlight the
most fundamental properties of SDS supported by the tightly localized gain
embedded into lossy optical media.

\subsection{Analytical considerations}

\subsubsection{Exact results}

Stationary solutions to Eq. (\ref{CGHS}) are looked for as $%
u(x,t)=e^{ikz}U(x),$ where complex function $U(x)$ satisfies an ordinary
differential equation,%
\begin{equation}
\left( \gamma +ik\right) U=\frac{i}{2}\frac{d^{2}U}{dx^{2}}-\left( \epsilon
-i\beta \right) |U|^{2}U,  \label{ODE}
\end{equation}%
at$~x\neq 0,$ supplemented by the boundary condition (b.c.) at $x=0$, which
is generated by the integration of Eq. (\ref{CGHS}) in an infinitesimal
vicinity of $x=0$,%
\begin{equation}
\lim_{x\rightarrow +0}\frac{d}{dx}U(x)=\left( i\Gamma _{1}-\Gamma
_{2}\right) U(x=0),  \label{bc}
\end{equation}%
assuming even stationary solutions, $U(-x)=U(x)$.

As seen from the expression for $A^{2}$ in Eq. (\ref{PSconstants}) [recall
that Eqs. (\ref{GL}) and (\ref{CGHS}) have opposite signs in front of $%
\gamma $], Eq. (\ref{ODE}) with $\gamma >0$ and $\epsilon >0$ cannot be
solved by a $\mathrm{sech}$ ansatz similar to that in Eq. (\ref{PS}). As an
alternative, $\mathrm{sech}$ can be replaced by $1/\sinh $:

\begin{equation}
U(x)=A\left[ \sinh \left( \kappa \left( |x|+\xi \right) \right) \right]
^{-\left( 1+i\mu \right) },  \label{sinh}
\end{equation}%
where $\xi >0$ prevents the singularity. This ansatz yields an exact \textit{%
codimension-one} solution to Eq. (\ref{ODE}) with b.c. (\ref{bc}), which is
valid under a special constraint imposed on coefficients of the system:%
\begin{equation}
\Gamma _{1}/\Gamma _{2}-2\Gamma _{2}/\Gamma _{1}=3\beta /\epsilon .
\label{codim}
\end{equation}%
Parameters of this solution are
\begin{equation}
A^{2}=\frac{3\gamma }{2\epsilon },~\kappa ^{2}=\gamma \frac{\Gamma _{2}}{%
\Gamma _{1}},~\mu =-\frac{\Gamma _{1}}{\Gamma _{2}},~k=\frac{\gamma }{2}%
\left( \frac{\Gamma _{2}}{\Gamma _{1}}-\frac{\Gamma _{1}}{\Gamma _{2}}%
\right) ,  \label{sol3}
\end{equation}%
\begin{equation}
\xi =\frac{1}{2}\sqrt{\frac{\Gamma _{1}}{\gamma \Gamma _{2}}}\ln \left(
\frac{\sqrt{\Gamma _{1}\Gamma _{2}}+\sqrt{\gamma }}{\sqrt{\Gamma _{1}\Gamma
_{2}}-\sqrt{\gamma }}\right) .  \label{xi}
\end{equation}%
The squared amplitude of the solution is%
\begin{equation}
\left\vert U(x=0)\right\vert ^{2}=\left( 3/2\epsilon \right) \left( \Gamma
_{1}\Gamma _{2}-\gamma \right) .  \label{Ampl}
\end{equation}

The main characteristic of the localized beam is its total power,
\begin{equation}
P=\int_{-\infty }^{+\infty }\left\vert u(x)\right\vert ^{2}dx.  \label{P}
\end{equation}%
For the solution given by Eqs. (\ref{sinh})-(\ref{xi}),%
\begin{equation}
P=\left( 3/\epsilon \right) \sqrt{\Gamma _{1}}\left( \sqrt{\Gamma _{1}}-%
\sqrt{\gamma /\Gamma _{2}}\right) .  \label{Ptot}
\end{equation}%
Obviously, solution (\ref{sinh}) exists if it yields $\left\vert
U(x=0\right\vert ^{2}>0$ and $P>0$, i.e.,
\begin{equation}
\Gamma _{1}>\left( \Gamma _{1}\right) _{\mathrm{thr}}\equiv \gamma /\Gamma
_{2}.  \label{thr}
\end{equation}%
The meaning of threshold condition (\ref{thr}) is that, to support the
stable pinned soliton, the local gain ($\Gamma _{1}$) must be sufficiently
large in comparison with the background loss, $\gamma $. It is relevant to
mention that, according to Eq. (\ref{codim}), the exact solution given by
Eqs. (\ref{sinh}), (\ref{sol3}), and (\ref{xi}) emerges at threshold (\ref%
{thr}) in the SDF medium, with $\beta <0$, provided that $\gamma <\sqrt{2}%
\Gamma _{2}^{2}$. In the opposite case, $\gamma >\sqrt{2}\Gamma _{2}^{2}$,
the threshold is realized in the SF medium, with $\beta >0$.

\subsubsection{Exact results for $\protect\gamma =0$ (no linear background
loss)}

The above analytical solution admits a nontrivial limit for $\gamma
\rightarrow 0$, which implies that the local gain compensates only the
nonlinear loss, accounted for by term $\sim \epsilon $ in Eq. (\ref{CGHS}).
In this limit, the pinned state is weakly localized, instead of the
exponentially localized one (\ref{sinh}):

\begin{equation}
U_{\gamma =0}(x)=\sqrt{\frac{3}{2\epsilon }}\frac{\sqrt{\Gamma _{1}/\Gamma
_{2}}}{\left( \left\vert x\right\vert +\Gamma _{2}^{-1}\right) ^{1+i\mu }}%
~,~k=0,  \label{gamma=0}
\end{equation}%
with $\mu $ given by expression (\ref{sol3}) (an overall phase shift is
dropped here). Note that the existence of solution (\ref{gamma=0}) does not
require any threshold condition, unlike Eq. (\ref{thr}). This weakly
localized state is a physically meaningful one, as its total power (\ref{P})
converges, $P(\gamma =0)=3\Gamma _{1}/\epsilon $. Note that this power does
not depend on the local-potential strength, $\Gamma _{2}$, unlike the
generic expression (\ref{Ptot}).

\subsubsection{Perturbative results for the self-defocusing medium}

In the limit case when the loss and gain vanish, $\gamma =\epsilon =\Gamma
_{1}=0$, solution (\ref{sinh}) goes over into an exact one in the SDF medium
(with $\beta <0$) pinned by the attractive potential:

\begin{eqnarray}
U(x)=\frac{\sqrt{2k/|\beta |}}{\sinh \left( \sqrt{2k}\left( |x|+\xi
_{0}\right) \right) },  \label{U0} \\
\xi _{0}=\frac{1}{2\sqrt{2k}}\ln \left( \frac{\Gamma _{2}+\sqrt{2k}}{\Gamma
_{2}-\sqrt{2k}}\right) ,  \label{xi0} \\
\left\vert U(x=0)\right\vert ^{2}=|\beta |^{-1}\left( \Gamma _{2}-2k\right) ,
\label{Amp0}
\end{eqnarray}%
which in interval $0<k<(1/2)\Gamma _{2}^{2}$ of the propagation constant.
The total power of this solution is%
\begin{equation}
P_{0}=\left( 2/|\beta |\right) \left( \Gamma _{2}-\sqrt{2k}\right) .
\label{P0}
\end{equation}%
In the limit of $k=0$, when amplitude (\ref{Amp0}) and power (\ref{P0})
attain their maxima, the solution degenerates from the exponentially
localized into a weakly localized one [cf. Eq. (\ref{gamma=0})],%
\begin{equation}
U_{k=0}(x)=\frac{1}{\sqrt{|\beta |}\left( |x|+\Gamma _{2}^{-1}\right) },
\label{k=0}
\end{equation}%
whose total power converges, $P_{0}(k=0)=2\Gamma _{2}/|\beta |$, as per Eq. (%
\ref{P0}).

The exact solutions given by Eqs. (\ref{U0})-(\ref{k=0}), which are generic
ones in the conservative model [no spacial constraint, such as (\ref{codim}%
), is required], may be used to construct an approximate solution to the
full system of Eqs. (\ref{ODE}) and (\ref{bc}), assuming that the gain and
loss parameters, $\Gamma _{1},$ $\gamma $, and $\epsilon $, are all small.
To this end, one can use the balance equation for the total power:%
\begin{equation}
\frac{dP}{dz}=-2\gamma P-2\epsilon \int_{-\infty }^{+\infty }\left\vert
u(x)\right\vert ^{4}dx+2\Gamma _{1}\left\vert u\left( x=0\right) \right\vert
^{2}=0.  \label{balance}
\end{equation}%
The substitution, in the zero-order approximation, of solution (\ref{U0}), (%
\ref{xi0}) into Eq. (\ref{balance}) yields the gain strength which is
required to compensate the linear and nonlinear losses in the solution with
propagation constant $k$:%
\begin{equation}
\Gamma _{1}=\frac{2\gamma }{\Gamma _{2}+\sqrt{2k}}+\frac{2\epsilon }{3|\beta
|}\frac{\left( \Gamma _{2}-\sqrt{2k}\right) \left( \Gamma _{2}+2\sqrt{2k}%
\right) }{\Gamma _{2}+\sqrt{2k}}.  \label{Gamma1}
\end{equation}%
As follows from Eq. (\ref{Gamma1}), with the decrease of $k$ from the
largest possible value, $\left( 1/2\right) \Gamma _{2}^{2}$, to $0$, the
necessary gain increases from the minimum, which exactly coincides with
threshold (\ref{thr}), to the largest value at which the perturbative
treatment admits the existence of the stationary pinned mode,
\begin{equation}
\left( \Gamma _{1}\right) _{\max }=2\gamma /\Gamma _{2}+2\epsilon \Gamma
_{2}/\left( 3|\beta |\right) .  \label{Gmax}
\end{equation}%
The respective total power grows from $0$ to the above-mentioned maximum, $%
2\Gamma _{2}/|\beta |$.

It is expected that, at $\Gamma _{1}$ exceeding the limit value (\ref{Gmax})
admitted by the stationary mode, the solution becomes nonstationary, with
the pinned mode emitting radiation waves, which makes the effective loss
larger and thus balances $\Gamma _{1}-\left( \Gamma _{1}\right) _{\max }$.
However, this issue was not studied in detail.

The perturbative result clearly suggests that, in the lossy SDF medium with
the local gain, the pinned modes exist not only under the special condition (%
\ref{codim}), at which they are available in the exact form, but as fully
generic solutions too. Furthermore, the increase of the power with the gain
strength implies that the modes are, most plausibly, stable ones.

\subsubsection{Perturbative results for the self-focusing medium}

In the case of $\beta >0$, which corresponds to the SF sign of the cubic
nonlinearity, a commonly known exact solution for the pinned mode in the
absence of the loss and gain, $\gamma =\epsilon =\Gamma _{1}=0$, is

\begin{eqnarray}
U(x)=\sqrt{2k/|\beta |}\mathrm{sech}\left( \sqrt{2k}\left( |x|+\xi
_{0}\right) \right) ,  \label{U02} \\
\xi _{0}=\frac{1}{2\sqrt{2k}}\ln \left( \frac{\sqrt{2k}+\Gamma _{2}}{\sqrt{2k%
}-\Gamma _{2}}\right) ,  \label{xi02} \\
\left\vert U(x=0)\right\vert ^{2}=|\beta |^{-1}\left( 2k-\Gamma _{2}\right) ,
\label{Amp02}
\end{eqnarray}%
with the total power%
\begin{equation}
P_{0}=\left( 2/\beta \right) \left( \sqrt{2k}-\Gamma _{2}\right) ,
\label{P02}
\end{equation}%
which exists for propagation constants $k>\left( 1/2\right) \Gamma _{2}^{2}$%
, cf. Eqs. (\ref{U0})-(\ref{P0}). In this case, the power-balance condition (%
\ref{balance}) yields a result which is essentially different from its
counterpart (\ref{Gamma1}):%
\begin{equation}
\Gamma _{1}=\frac{2\gamma }{\sqrt{2k}+\Gamma _{2}}+\frac{2\epsilon }{3\beta }%
\frac{\left( \sqrt{2k}-\Gamma _{2}\right) \left( 2\sqrt{2k}+\Gamma
_{2}\right) }{\sqrt{2k}+\Gamma _{2}}.  \label{Gamma12}
\end{equation}%
Straightforward consideration of Eq. (\ref{Gamma12}) reveals the difference
of the situation from that considered above for the SDF medium: if the
strength of the nonlinear loss is relatively small,%
\begin{equation}
\epsilon <\epsilon _{\mathrm{cr}}=\left( \beta /2\Gamma _{2}^{2}\right)
\gamma ,  \label{epscr}
\end{equation}%
the growth of power (\ref{P02}) from zero at $\sqrt{2k}=\Gamma _{2}$ with
the increase of $k$ is initially (at small values of $P_{0}$) requires not%
\emph{\ }the increase of the gain strength from the threshold value (\ref%
{thr}) to $\Gamma _{1}>\left( \Gamma _{1}\right) _{\mathrm{thr}}$, but, on
the contrary, \emph{decrease} of $\Gamma _{1}$ to $\Gamma _{1}<\left( \Gamma
_{1}\right) _{\mathrm{thr}}$ \cite{Mak}. Only at $\epsilon >\epsilon _{%
\mathrm{cr}}$, see Eq. (\ref{epscr}), the power grows with $\Gamma _{1}$
starting from $\Gamma _{1}=\left( \Gamma _{1}\right) _{\mathrm{thr}}$.

\subsubsection{The stability of the zero solution, and its relation to the
existence of pinned solitons}

It is possible to check the stability of the zero solution, which is an
obvious prerequisite for the soliton's stability, as said above. To this
end, one should use the linearized version of Eq. (\ref{CGHS}),%
\begin{equation}
\frac{\partial u_{\mathrm{lin}}}{\partial z}=\frac{i}{2}\frac{\partial
^{2}u_{\mathrm{lin}}}{\partial x^{2}}-\gamma u_{\mathrm{lin}}+\left( \Gamma
_{1}+i\Gamma _{2}\right) \delta (x)u_{\mathrm{lin}}.  \label{linearized}
\end{equation}%
The critical role is played by localized eigenmodes of Eq. (\ref{linearized}%
),
\begin{equation}
u_{\mathrm{lin}}(x,t)=u_{0}e^{\Lambda z}e^{il\left\vert x\right\vert
-\lambda \left\vert x\right\vert },  \label{mode}
\end{equation}%
where $u_{0}$ is an arbitrary amplitude, localization parameter $\lambda $
must be positive, $l$ is a wavenumber, and $\Lambda $ is a complex
instability growth rate. Straightforward analysis yields \cite{HSexact}
\begin{equation}
\lambda -il=\Gamma _{2}-i\Gamma _{1},~\Lambda =(i/2)\left( \Gamma
_{2}-i\Gamma _{1}\right) ^{2}-\gamma .  \label{lambda}
\end{equation}%
It follows from Eq. (\ref{lambda}) that the stability condition for the zero
solution, $\mathrm{Re}\left( \Lambda \right) <0$, amounts to inequality $%
\gamma >\Gamma _{1}\Gamma _{2},$ which is exactly \emph{opposite} to Eq. (%
\ref{thr}). In fact, Eq. (\ref{Ampl}) demonstrates that the exact
pinned-soliton solution given by Eqs. (\ref{sinh})-(\ref{xi}) emerges, via
the the standard \textit{forward} (alias \textit{supercritical}) \textit{%
pitchfork bifurcation} \cite{bif}, precisely at the point where the zero
solution loses its stability to the local perturbation. The above analysis
of Eq. (\ref{Gamma1}) demonstrates that the same happens with the
perturbative solution (\ref{U0}), (\ref{xi0}) in the SDF model. On the
contrary, the analysis of Eq. (\ref{Gamma12}) has revealed above that the
same transition happens to the perturbative solution (\ref{U02}), (\ref{xi02}%
) in the SF medium only at $\epsilon >\epsilon _{\mathrm{cr}}$, see Eq. (\ref%
{epscr}), while at $\epsilon >\epsilon _{\mathrm{cr}}$ the pitchfork
bifurcation is of the \textit{backward} (alias \textit{subcritical} \cite%
{bif}) type, featuring the power which originally grows with the \emph{%
decrease} of the gain strength. Accordingly, the pinned modes emerging from
the subcritical bifurcation are unstable. However, the contribution of the
nonlinear loss ($\epsilon >0$) eventually leads to the turn of the solution
branch forward and its stabilization at the turning point. For very small $%
\epsilon $, the turning point determined by Eq. (\ref{Gamma12}) is located
at $\Gamma _{1}\approx 4\sqrt{2\gamma \epsilon /\left( 3\beta \right) }$, $%
P_{0}\approx \sqrt{6\gamma /\left( \beta \epsilon \right) }$.

Lastly, note that the zero solution is never destabilized, and the stable
pinned soliton does \ not emerge, in the absence of the local attractive
potential, i.e., at $\Gamma _{2}\leq 0$.

\subsection{Numerical results}

\subsubsection{Self-trapping and stability of the pinned solitons}

\begin{figure}[tbp]
\includegraphics[width=7cm]{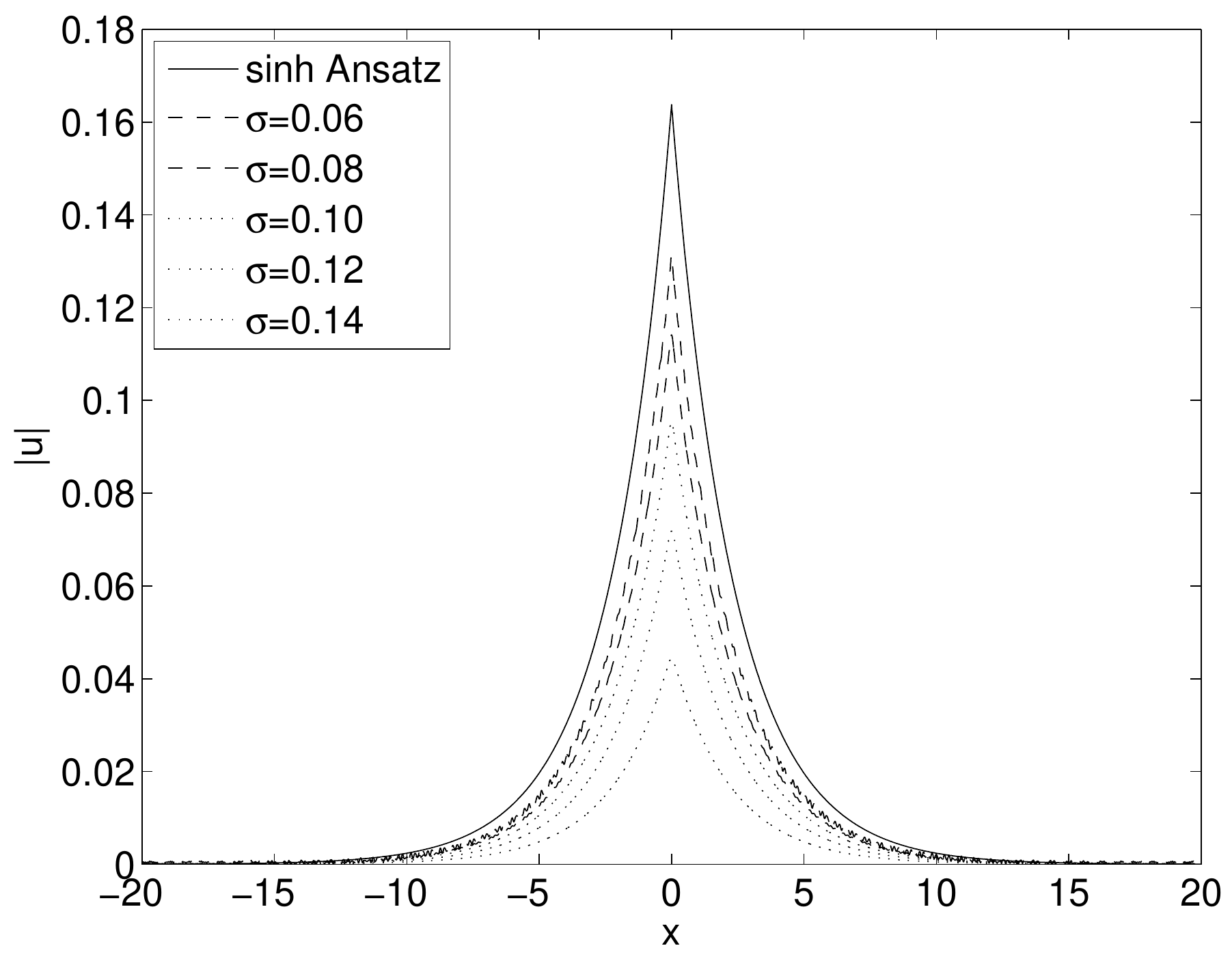}
\caption{The exact solution for the pinned soliton given by Eqs. (\protect
\ref{sinh})-(\protect\ref{Ampl}), and a set of approximations generated by
the regularized delta-function defined as per Eq. (\protect\ref{delta}). All
the profiles represent stable solutions. Other parameters are $\protect\beta %
=0,~$ $\protect\gamma =0.25,$ $\Gamma _{1}=0.6155$, and $\Gamma _{2}=\Gamma
_{1}/\protect\sqrt{2}$, as per Eq. (\protect\ref{codim}).}
\label{fig1}
\end{figure}

The numerical analysis of the model based on Eq. (\ref{CGHS}) was performed
with the delta-function replaced by its Gaussian approximation,%
\begin{equation}
\tilde{\delta}(x)=\left( \sqrt{\pi }\sigma \right) ^{-1}\exp \left(
-x^{2}/\sigma ^{2}\right) ,  \label{delta}
\end{equation}%
with finite width $\sigma $. The shape of a typical analytical solution (\ref%
{sinh})-(\ref{Ampl} for the pinned soliton, and a set of approximations to
it provided by the use of approximation (\ref{delta}), are displayed in Fig. %
\ref{fig1}. All these pinned states are stable, as was checked by
simulations of their perturbed evolution in the framework of Eq. (\ref{CGHS}%
) with $\delta (x)$ replaced by $\tilde{\delta}(x)$.

The minimum (threshold) value of the local-gain strength, $\Gamma _{1}$,
which is necessary for the existence of stable pinned solitons is an
important characteristic of the setting, see Eq. (\ref{thr}). Figure \ref%
{fig2} displays the dependence of the $\left( \Gamma _{1}\right) _{\mathrm{%
thr}}$ on the strength of the background loss, characterized by $\sqrt{%
\gamma }$, for $\beta =0$ and three fixed values of the local-potential's
strength, $\Gamma _{2}=+1,0,-1$ (in fact, the solutions corresponding to $%
\Gamma _{2}=-1$ are unstable, as they are repelled by the HS). In addition, $%
\left( \Gamma _{1}\right) _{\mathrm{thr}}$ is also shown as a function of $%
\sqrt{\gamma }$ under constraint (\ref{codim}), which amounts $\Gamma
_{2}=\Gamma _{1}/\sqrt{2}$, in the case of $\beta =0$. The corresponding
analytical prediction, as given by Eq. (\ref{thr}), is virtually identical
to its numerical counterpart, despite the difference of approximation (\ref%
{delta}) from the ideal delta-function.
\begin{figure}[tbp]
\includegraphics[width=7cm]{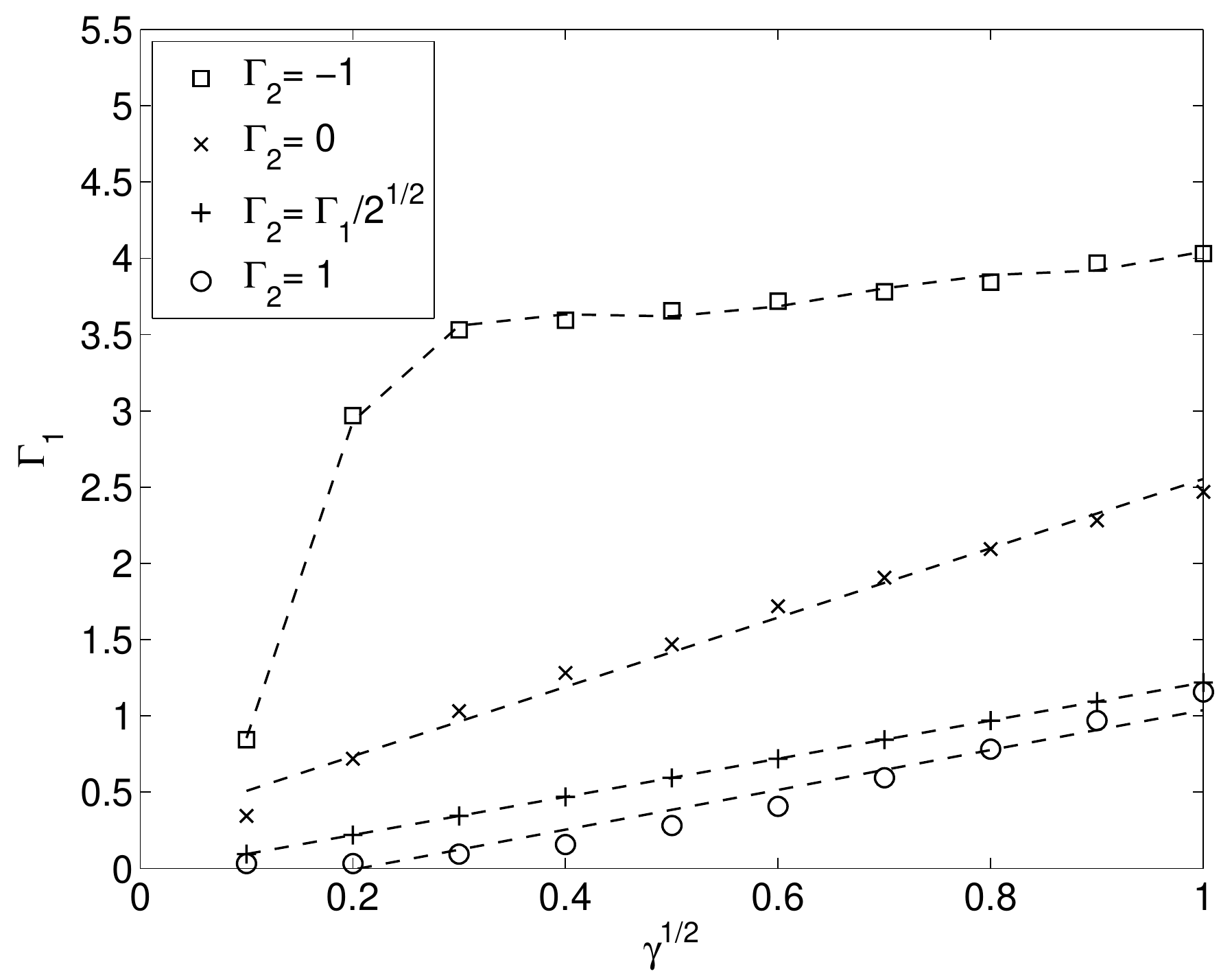}
\caption{Chains of symbols show the minimum value of the local gain, $\Gamma
_{1}$, which is necessary for the creation of stationary pinned solitons in
the framework of Eq. (\protect\ref{CGHS}), as a function of the background
loss ($\protect\gamma $), with $\protect\beta =0,$ $\protect\varepsilon =1$,
and $\protect\delta (x)$ approximated as per Eq. (\protect\ref{delta}) with $%
\protect\sigma =0.1$. The strength of the local potential, $\Gamma _{2}$, is
fixed as indicated in the box. For $\Gamma _{2}=\pm 1$ and $0,$ the lines
are guides for the eye, while the straight line for the case of $\Gamma
_{2}=\Gamma _{1}/\protect\sqrt{2}$, which corresponds to Eq. (\protect\ref%
{codim}) with $\protect\beta =0$, is the analytical prediction given by Eq. (%
\protect\ref{thr}). }
\label{fig2}
\end{figure}

Figure \ref{fig2} corroborates that, as said above, the analytical solutions
represent only a particular case of the family of generic dissipative
solitons that can be found in the numerical form. In particular, the
solution produces narrow and tall stable pinned solitons at large values of $%
\Gamma _{1}$. All the pinned solitons, including weakly localized ones
predicted by analytical solution (\ref{gamma=0}) for $\gamma =0$, are stable
at $\Gamma _{1}>\left( \Gamma _{1}\right) _{\mathrm{thr}}$ and $\Gamma
_{2}>0 $.

\begin{figure}[tbp]
\includegraphics[width=8cm]{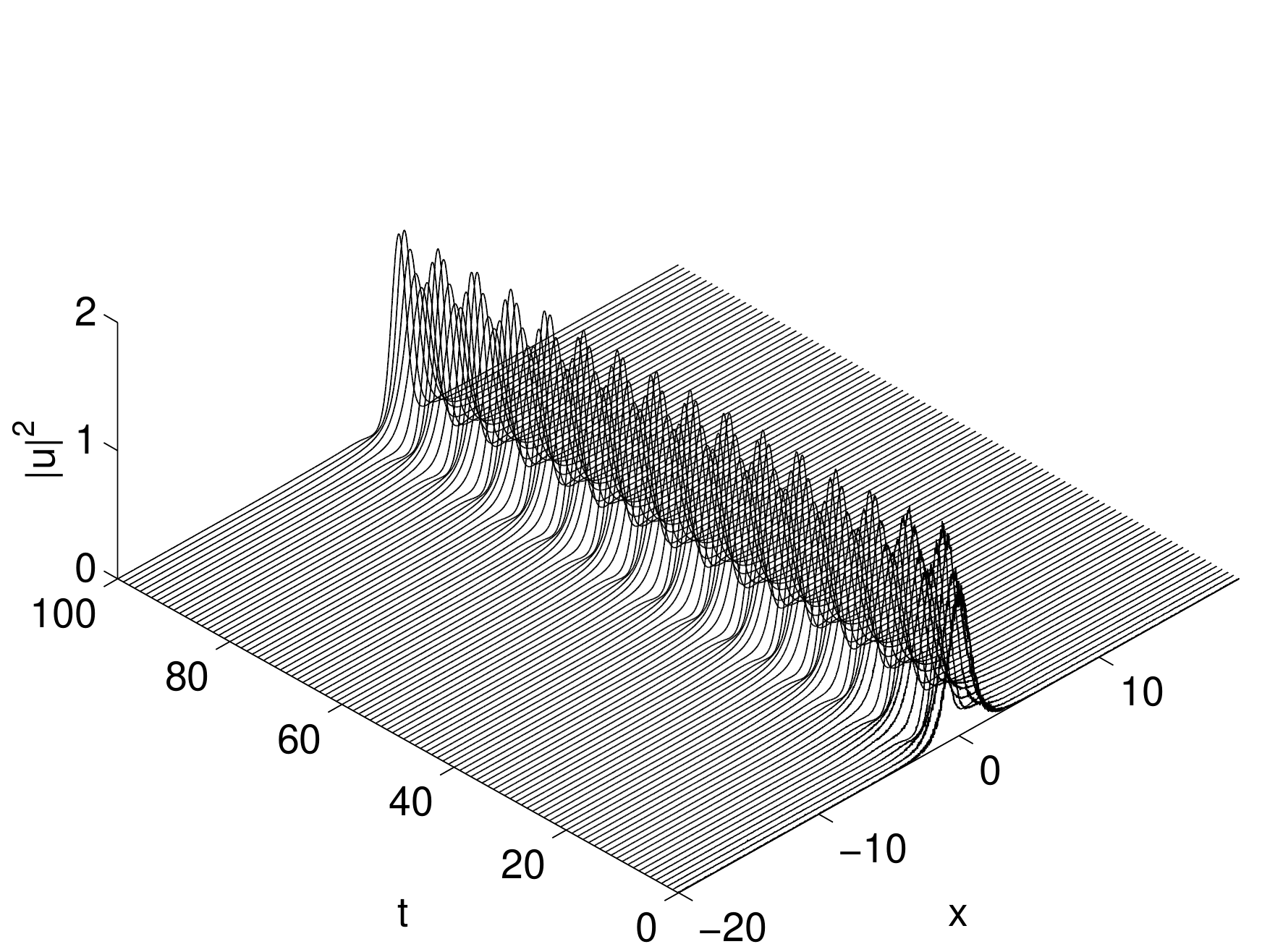}
\caption{A typical example of a robust breather produced by
simulations of Eq. (\protect\ref{CGHS}) with $\protect\delta (x)$
replaced by approximation
$\tilde{\protect\delta}(x)$ as per Eq. (\protect\ref{delta}) with $\protect%
\sigma =2$, the other parameters being $\protect\beta =\protect\epsilon %
=\Gamma _{2}=1$, $\Gamma _{1}=4$, and $\protect\gamma =0.1$.}
\label{fig3}
\end{figure}

The situation is essentially different for large $\sigma $ in Eq. (\ref%
{delta}), i.e., when the local gain is supplied in a broad region.
In that case, simulations do not demonstrate self-trapping into
stationary solitons; instead, a generic outcome is the formation of
stable \textit{breathers} featuring regular intrinsic oscillations,
the breather's width being on the same order of magnitude as $\sigma
$, see a typical example in Fig. \ref{fig3}. 
It seems plausible that, with the increase of
$\sigma $, the static pinned soliton is destabilized via the Hopf
bifurcation \cite{bif} which gives rise to the stable breather.

\subsection{The model with the double hot spot}

\begin{figure}[tbp]
\includegraphics[width=6cm]{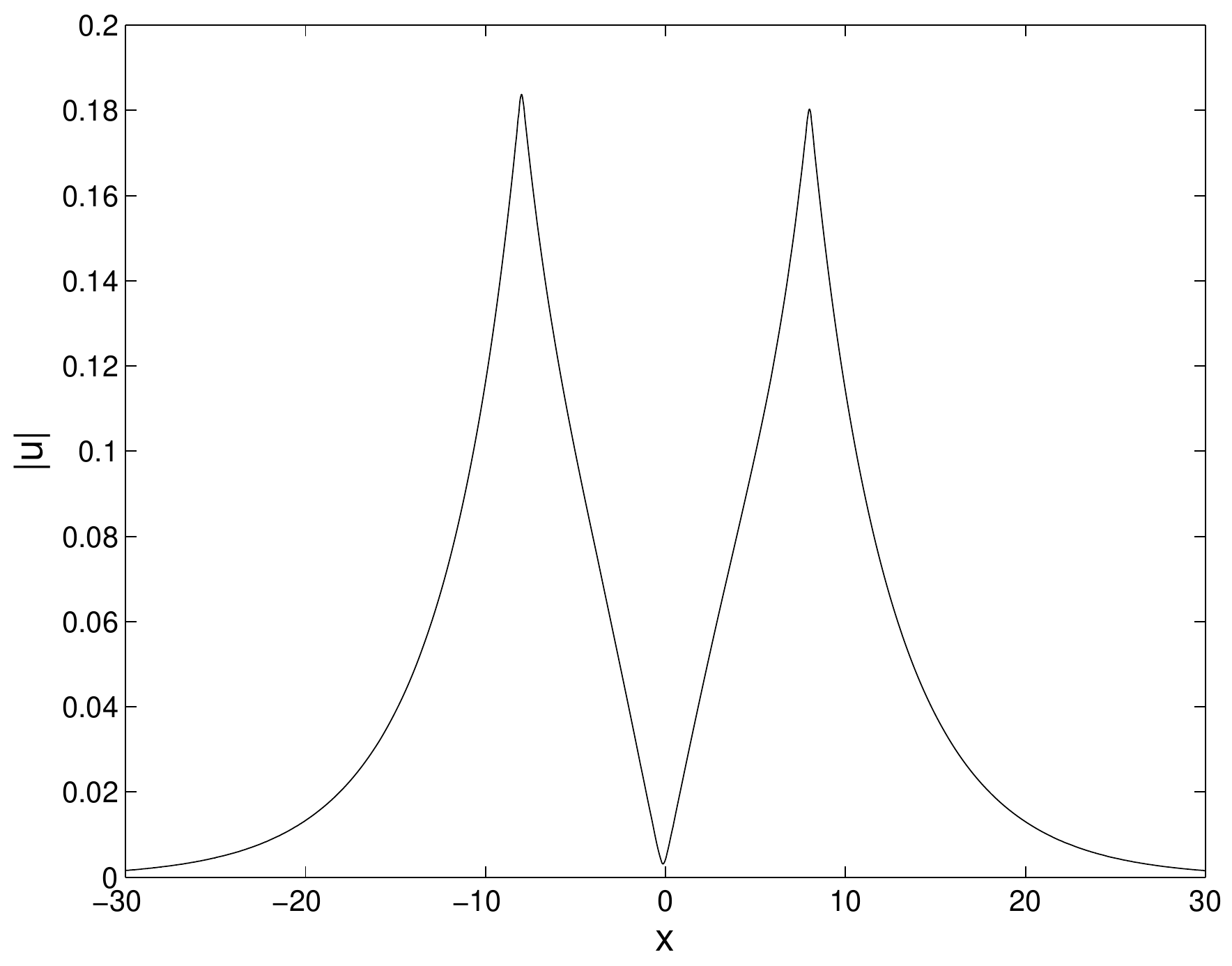}
\caption{A stable symmetric mode generated by Eq. (\protect\ref{two1}) with
the delta-functions approximated as per Eq. (\protect\ref{delta}) with $%
\protect\sigma =0.1$. Other parameters are $\protect\beta =0,\protect%
\epsilon =1,\protect\gamma =0.057,\Gamma _{1}=0.334$, $\Gamma _{2}=0.236$,
and $L=8$.}
\label{fig4}
\end{figure}

The extension of Eq. (\ref{CGHS}) for two mutually symmetric HSs separated
by distance $2L$ was introduced in \cite{spotsExact1}:

\begin{eqnarray}
\frac{\partial u}{\partial z}=\frac{i}{2}\frac{\partial ^{2}u}{\partial x^{2}%
}-\gamma u-\left( \epsilon -i\beta \right) |u|^{2}u  \notag \\
+\left( \Gamma _{1}+i\Gamma _{2}\right) \left[ \delta \left( x+L\right)
+\delta \left( x-L\right) \right] u.  \label{two1}
\end{eqnarray}%
Numerical analysis has demonstrated that stationary symmetric solutions of
this equation (they are not available in an analytical form) are stable, see
a typical example in Fig. \ref{fig4}, while\ all antisymmetric states are
unstable, spontaneously transforming into their symmetric counterparts.

\begin{figure}[tbp]
\includegraphics[width=8cm]{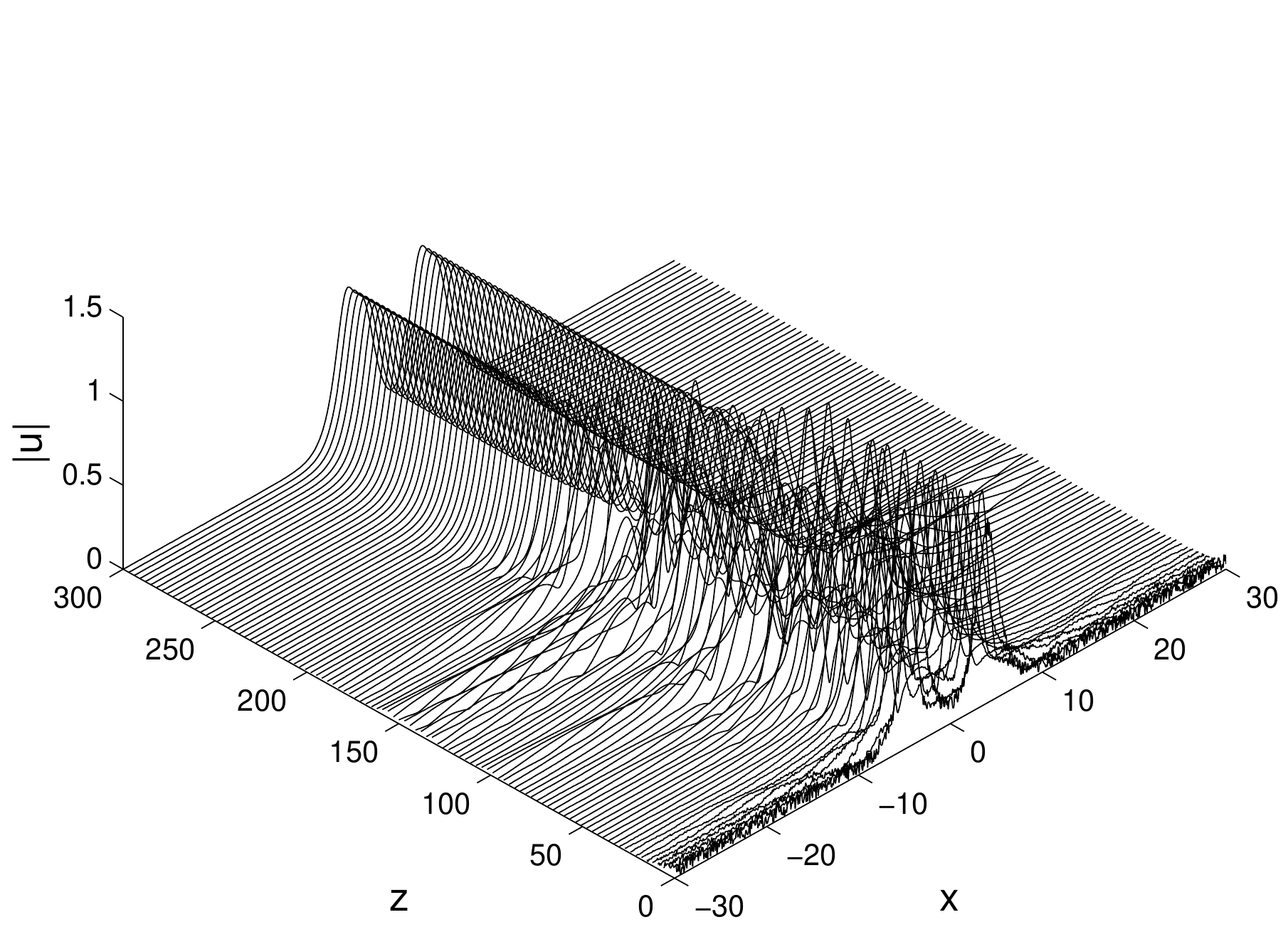}
\caption{Spontaneous transformation of a pair of breathers pinned to two hot
spots, described by Eq. (\protect\ref{two1}), into a stationary symmetric
mode. Parameters are $\protect\beta =0$, $\protect\epsilon =1$, $\protect%
\gamma =1$, $\Gamma _{1}=4$, $\Gamma _{2}=1$, and $L=4$.}
\label{fig5}
\end{figure}

As shown above in Fig. \ref{fig3}, Eq. (\ref{CGHS}) with the single HS
described by expression (\ref{delta}), where $\sigma $ is large enough,
supports breathers, instead of stationary pinned modes. Simulations of Eq. (%
\ref{two1}) demonstrate that a pair of such broad HSs support unsynchronized
breathers pinned by each HS, if the distance between them is large enough,
hence the breathers virtually do not interact. A completely different effect
is displayed in Fig. \ref{fig5}, for two broad HSs which are set closer to
each other: the interaction between the breathers pinned to each HS
transforms them into a \emph{stationary} stable symmetric mode. It is
relevant to stress that the transformation of the breather pinned by the
isolated HS into a stationary pinned state does not occur for the same
parameters. This outcome is a generic result of the interaction between the
breathers, provided that the distance between them is not too large.

\subsection{Related models}

In addition to the HSs embedded into the medium with the uniform
nonlinearity, more specific models, in which the nonlinearity is also
concentrated at the HSs, were introduced in \cite{spotsExact1} and \cite%
{spotsExact2}. In particular, the respective system with the double HS is
described by the following equation:%
\begin{eqnarray}
\frac{\partial u}{\partial z}=\frac{i}{2}\frac{\partial ^{2}u}{\partial x^{2}%
}-\gamma u  \notag \\
+\left[ \left( \Gamma _{1}+i\Gamma _{2}\right) -\left( E-iB\right)
\left\vert u\right\vert ^{2}\right] \left[ \delta \left( x+L\right) +\delta
\left( x-L\right) \right] u,  \label{ad-hoc}
\end{eqnarray}%
where $B$ and $E$ are coefficients of the localized Kerr nonlinearity and
cubic loss, respectively. These settings may be realized if the nonlinear
properties of the waveguides are dominated by narrow doped segments.

Although model (\ref{ad-hoc}) seems somewhat artificial, its advantage is a
possibility to find both symmetric and antisymmetric pinned modes in an
exact analytical form. Those include both fundamental modes, with exactly
two local power peaks, tacked to the HSs, and higher-order states, which
feature additional peaks between the HSs. An essential finding pertains to
the stability of such states, for which the sign of the cubic nonlinearity
plays a crucial role: in the SF case [$B>0$ in Eq. (\ref{ad-hoc})], only the
fundamental symmetric and antisymmetric modes, with two local peaks tacked
to the HSs, may be stable. In this case, all the higher-order multi-peak
modes, being unstable, evolve into the fundamental ones. In the case of the
SDF cubic nonlinearity [$B<0$ in Eq. (\ref{ad-hoc})], the HS pair gives rise
to \emph{multistability}, with up to eight coexisting stable multi-peak
patterns, both symmetric and antisymmetric ones. The system without the Kerr
term ($B=0$), the nonlinearity in Eq. (\ref{ad-hoc}) being represented
solely by the local cubic loss ($\sim E$) is similar to one with the
self-focusing or defocusing nonlinearity, if the linear potential of the HS
is, respectively, attractive or repulsive, i.e., $\Gamma _{2}>0$ or $\Gamma
_{2}<0$ (note that a set of two local repulsive potentials may stably trap
solitons in an effective \textit{cavity} between them \cite{Mak}). An
additional noteworthy feature of the former setting is the coexistence of
the stable fundamental modes with robust breathers.

\section{Solitons pinned to the $\mathcal{PT}$-symmetric dipole}

\subsection{Analytical results}

The nonlinear Schr\"{o}dinger equation (NLSE) in the form of Eq. (\ref{eq})
is a unique example of a $\mathcal{PT}$-symmetric system in which a full
family of solitons can be found in an exact analytical form \cite{Thawatchai}%
. Indeed, looking for stationary solutions with real propagation constant $k$
as $u\left( x,z\right) =e^{ikz}U(x)$, where the $\mathcal{PT}$ symmetry is
provided by condition $U^{\ast }(x)=U(-x)$, one can readily find, for the SF
and SDF signs of the nonlinearity, respectively,
\begin{eqnarray}
U(x) &=&\sqrt{2k}\frac{\cos \theta +i~\mathrm{sgn}(x)\sin \theta }{\cosh
\left( \sqrt{2k}\left( |x|+\xi \right) \right) },\mathrm{~for}~~\sigma =+1,
\label{SF} \\
U(x) &=&\sqrt{2k}\frac{\cos \theta +i~\mathrm{sgn}(x)\sin \theta }{\sinh
\left( \sqrt{2k}\left( |x|+\xi \right) \right) },~\mathrm{for}~~\sigma =-1,
\label{SDF}
\end{eqnarray}%
with real constants $\theta $ and $\xi $, cf. Eq. (\ref{sinh}). The form of
this solution implies that $\mathrm{Im}\left( U(x=0)\right) =0,$ while jumps
($\Delta $) of the imaginary part and first derivative of the real part at $%
x=0$ are determined by the b.c. produced the integration of the $\delta $-
and $\delta ^{\prime }$- functions in an infinitesimal vicinity of $x=0$,
cf. Eq. (\ref{bc}):%
\begin{eqnarray}
\Delta \left\{ \mathrm{Im}\left( U\right) \right\} |_{x=0} &=&2\gamma
\mathrm{Re}\left( U\right) |_{x=0},  \label{Im} \\
\Delta \left\{ \left( \frac{d}{dx}\mathrm{Re}\left( U\right) \right)
\right\} |_{x=0} &=&-2\varepsilon _{0}~\mathrm{Re}\left( U\right) |_{x=0}.
\label{Re}
\end{eqnarray}%
The substitution of expressions (\ref{SF}) and (\ref{SDF}) into these b.c.
yields%
\begin{equation}
\theta =\arctan \left( \gamma \right) ,  \label{theta}
\end{equation}%
which does not depend on $k$ and is the same for $\sigma =\pm 1$, and%
\begin{equation}
\xi =\frac{1}{2\sqrt{2k}}\ln \left( \sigma \frac{\sqrt{2k}+\varepsilon _{0}}{%
\sqrt{2k}-\varepsilon _{0}}\right) ,  \label{xi2}
\end{equation}%
which does not depend on the $\mathcal{PT}$ coefficient, $\gamma $.

The total power (\ref{P}) of the localized mode is
\begin{equation}
P_{\sigma }=2\sigma \left( \sqrt{2k}-\varepsilon _{0}\right) .  \label{Power}
\end{equation}%
As seen from Eq. (\ref{xi2}), the solutions exist at
\begin{equation}
\left\{
\begin{array}{cc}
\sqrt{2k}>\varepsilon _{0} & \mathrm{for}~~\sigma =+1, \\
\sqrt{2k}<\varepsilon _{0} & \mathrm{for~~}\sigma =-1.%
\end{array}%
\right. ~~  \label{><}
\end{equation}%
As concerns stability of the solutions, it is relevant to mention that
expression (\ref{Power}) with $\sigma =+1$ and $-1$ satisfy, respectively,
the Vakhitov-Kolokolov (VK) \cite{VK,Berge} and ``anti-VK" \cite{anti}
criteria, i.e.,
\begin{equation}
dP_{+1}/dk>0,~dP_{-1}/dk<0,  \label{Vakh}
\end{equation}%
which are necessary conditions for the stability of localized modes
supported, severally, by the SF and SDF nonlinearities, hence both solutions
have a chance to be stable.

\subsection{Numerical findings}

As above [see Eq. (\ref{delta})], the numerical analysis of the model needs
to replace the exact $\delta $-function by its finite-width regularization, $%
\tilde{\delta}(x)$. In the present context, the use of the Gaussian
approximation is not convenient, as, replacing the exact solutions in the
form of Eqs. (\ref{SF}) and (\ref{SDF}) by their regularized counterparts,
it is necessary, \textit{inter alia}, to replace $\mathrm{sgn}(x)\equiv
-1+2\int_{-\infty }^{x}\delta (x^{\prime })dx^{\prime }$ by a continuous
function realized as $-1+2\int_{-\infty }^{x}\tilde{\delta}(x^{\prime
})dx^{\prime }$, which would be a non-elementary function. Therefore, the
regularization was used in the form of the Lorentzian,%
\begin{eqnarray}
\delta (x)\rightarrow \frac{a}{\pi }\frac{1}{x^{2}+a^{2}},~\delta ^{\prime
}(x)\rightarrow -\frac{2a}{\pi }\frac{x}{\left( x^{2}+a^{2}\right) ^{2}},~
\notag \\
\mathrm{sgn}(x)\rightarrow \frac{2}{\pi }\arctan \left( \frac{x}{a}\right) ,
\label{Lorentz}
\end{eqnarray}%
with $0<a\ll k^{-1/2}$ \cite{Thawatchai}.

\begin{figure}[tbp]
\centering$%
\begin{array}{c}
\includegraphics[width=2.5in]{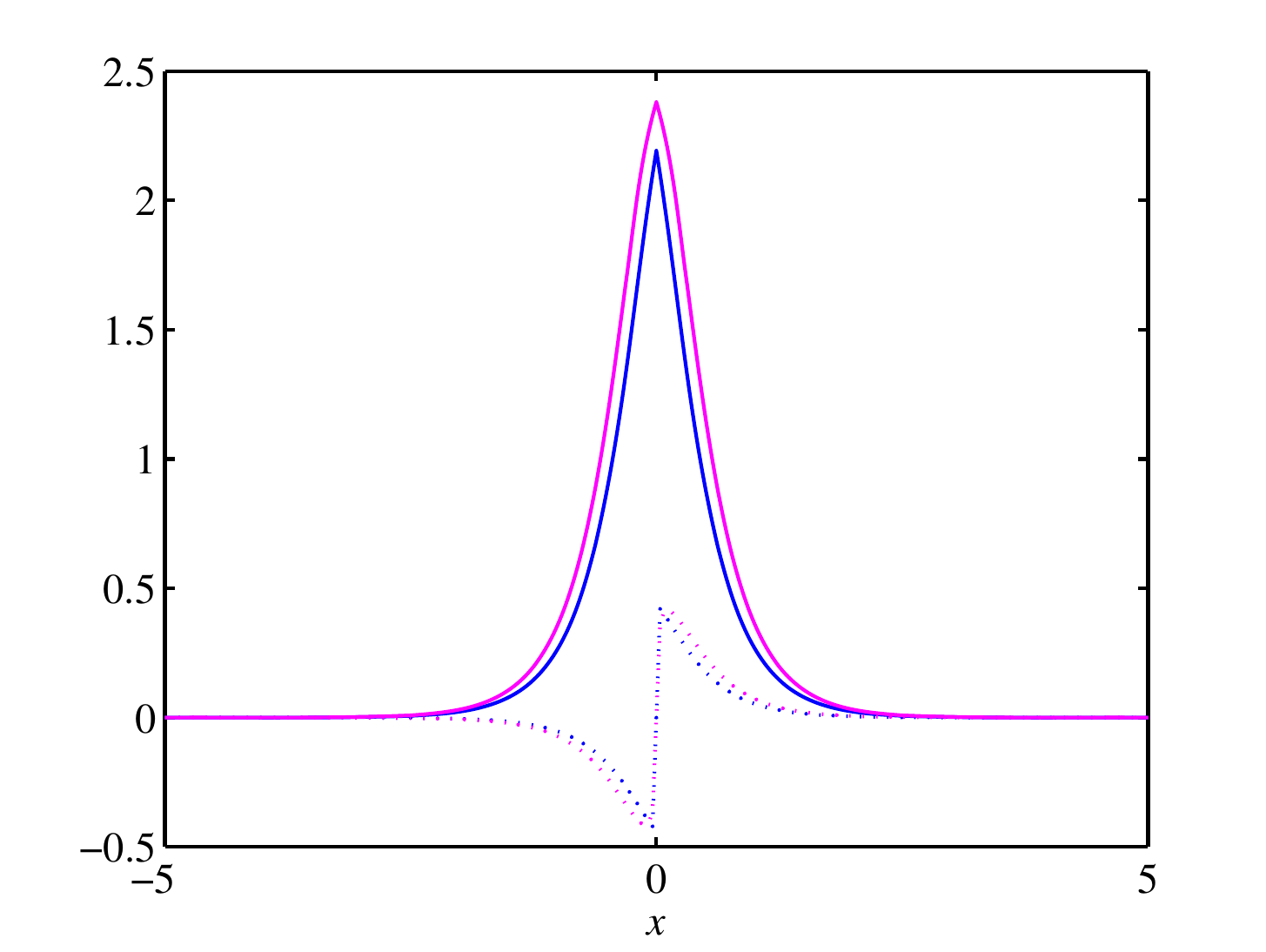} \\
(\mathrm{a}) \\
\includegraphics[width=2.5in]{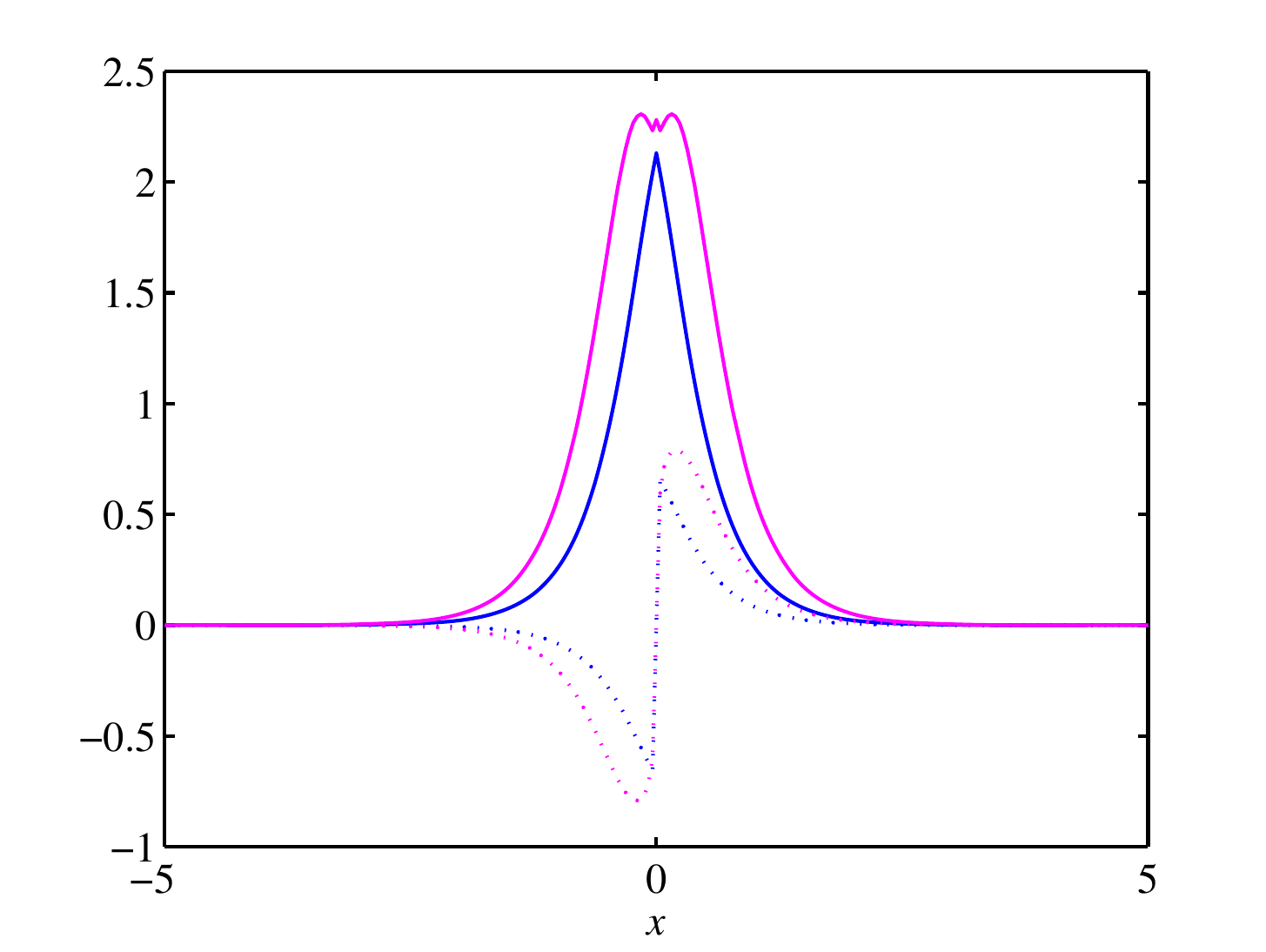} \\
(\mathrm{b})%
\end{array}%
$%
\caption{(Color online) Comparison between the analytical solutions (solid
and dotted blue curves show their real and imaginary parts, respectively),
given by Eqs. (\protect\ref{SF}), (\protect\ref{theta}), and (\protect\ref%
{xi2}) with $\protect\sigma =+1$ and $\protect\varepsilon _{0}=1$, and their
numerically found counterparts, obtained by means of regularization (\protect
\ref{Lorentz}) with $a=0.02$ (magenta curves). The $\mathcal{PT}$ parameter
is $\protect\gamma =0.20$ in (a) and $0.32$ in (b). In both panels, the
solutions are produced for propagation constant $k=3$.}
\label{fig6}
\end{figure}

The first result for the SF nonlinearity, with $\sigma =+1$ (in this case, $%
\varepsilon _{0}=+1$ is also fixed by scaling) is that, for given $a$ in Eq.
(\ref{Lorentz}), there is a critical value, $\gamma _{\mathrm{cr}}$, of the $%
\mathcal{PT}$ coefficient, such that at $\gamma <\gamma _{\mathrm{cr}}$ the
numerical solution features a shape very close to that of the analytical
solution (\ref{SF}), while at $\gamma >\gamma _{\mathrm{cr}}$ the
single-peak shape of the solution transforms into a \emph{double-peak} one,
as shown in Fig. \ref{fig6}(a). In particular, $\gamma _{\mathrm{cr}}\left(
a=0.02\right) \approx 0.24$.

\begin{figure}[tbp]
\centering\includegraphics[width=3in]{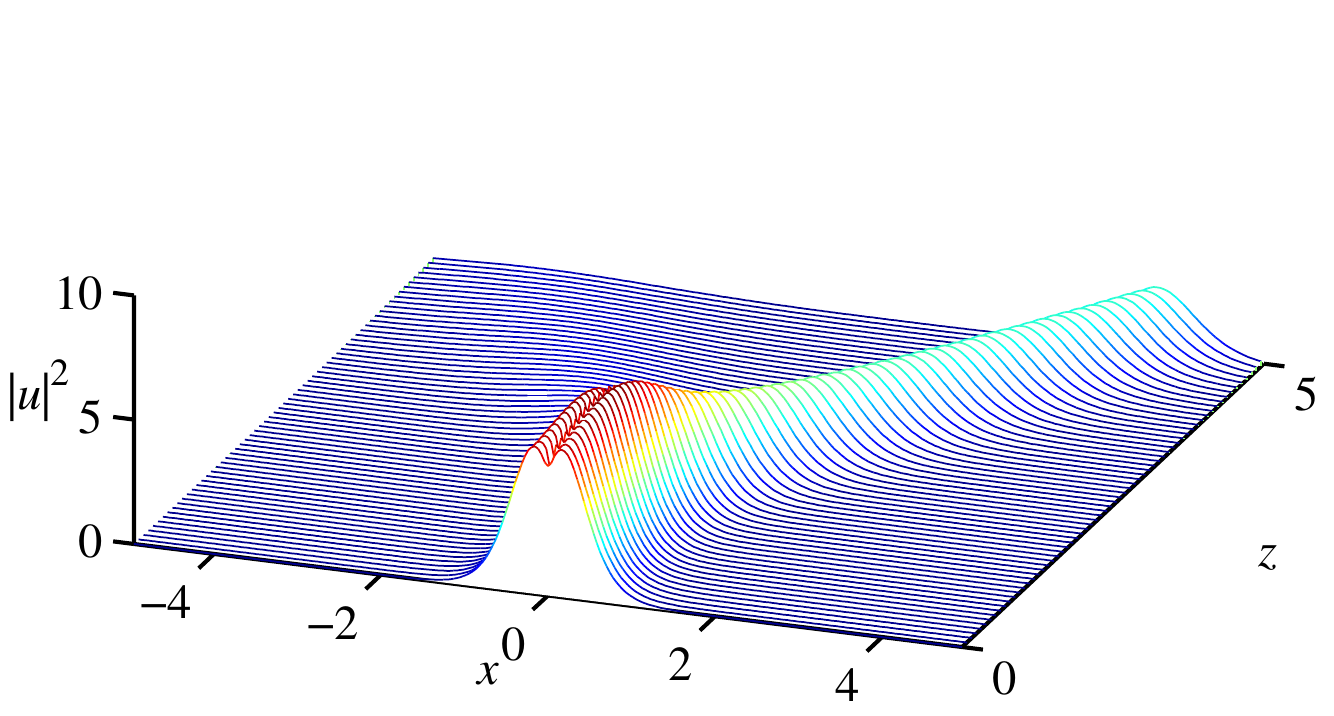} \caption{(Color
online) The unstable evolution (spontaneous escape) of the
double-peak soliton whose stationary form is shown in Fig. \protect\ref{fig6}%
(b).}
\label{fig7}
\end{figure}

The difference between the single- and double-peak modes is that the former
ones are completely stable, as was verified by simulations of Eq. (\ref{eq})
with regularization (\ref{Lorentz}), while all the double-peak solutions are
unstable. This correlation between the shape and (in)stability of the pinned
modes is not surprising: the single- and double-peak structures imply that
the pinned mode is feeling, respectively, effective attraction to or
repulsion from the local defect. Accordingly, in the latter case the pinned
soliton is unstable against spontaneous escape, transforming itself into an
ordinary freely moving NLSE soliton, as shown in Fig. \ref{fig7}.

\begin{figure}[tbp]
\centering\includegraphics[width=3in]{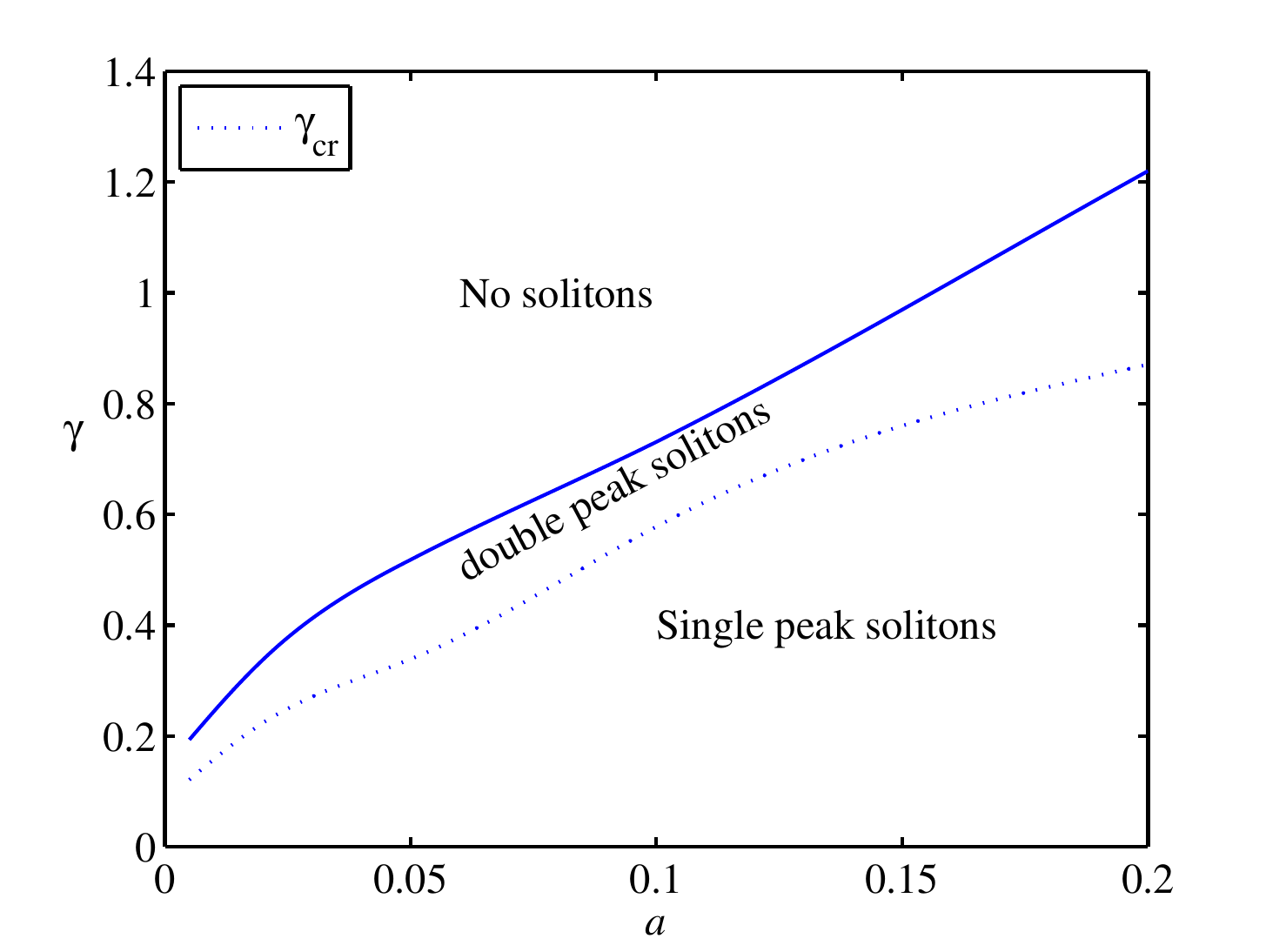} \caption{(Color
online) Regions of the existence of stable single-peak and
unstable double-peak $\mathcal{PT}$-symmetric solitons, separated by $%
\protect\gamma =\protect\gamma _{\mathrm{cr}}(a)$, in the plane of the
regularization scale, $a$, and the gain-loss parameter, $\protect\gamma ,$
for fixed $k=3.0$, in the system described by Eqs. (\protect\ref{eq}) and (%
\protect\ref{Lorentz}) with $\protect\sigma =+1$ (the SF nonlinearity) and $%
\protect\varepsilon _{0}=1$.}
\label{fig8}
\end{figure}

Figure \ref{fig8} summarizes the findings in the plane of $\left( a,\gamma
\right) $ for a fixed propagation constant, $k=3.0$. The region of the
unstable double-peak solitons is a relatively narrow boundary layer between
broad areas in which the stable single-peak solitons exist, as predicted by
the analytical solution, or no solitons exist at all, at large values of $%
\gamma $. Note also that the stability area strongly expands to larger
values of $\gamma $ as the regularized profile (\ref{Lorentz}) becomes
smoother, with the increase of $a$. On the other hand, the stability region
does not vanish even for very small $a$.
For the same model but with the SDF nonlinearity, $\sigma =-1$, the results
are simpler: all the numerically found pinned solitons are close to the
analytical solution (\ref{SDF}), featuring a single-peak form, and are
completely stable.

A generalization of model (\ref{eq}), including a nonlinear part of the
trapping potential, was studied in \cite{Thawatchai} too:%
\begin{equation}
i\frac{\partial u}{\partial z}=-\frac{1}{2}\frac{\partial ^{2}u}{\partial
x^{2}}-\sigma |u|^{2}u-\left( \varepsilon _{0}+\varepsilon
_{2}|u|^{2}\right) u\delta \left( x\right) +i\gamma u\delta ^{\prime }\left(
x\right) .  \label{02}
\end{equation}%
Equation (\ref{02}) also admits an analytical solution for pinned solitons,
which is rather cumbersome. It takes a simpler form in the case when the
trapping potential at $x=0$ is purely nonlinear, with $\varepsilon _{0}=0$, $%
\varepsilon _{2}>0$, Eq. (\ref{xi2}) being replaced by%
\begin{equation}
\xi =\frac{1}{2\sqrt{2k}}\ln \left[ \frac{2\varepsilon _{2}\sqrt{2k}}{%
1+\gamma ^{2}}+\sqrt{1+\frac{8\varepsilon _{2}^{2}k}{\left( 1+\gamma
^{2}\right) ^{2}}}\right]  \label{xi3}
\end{equation}%
[it is the same for both signs of the bulk nonlinearity, $\sigma =+1$ and $%
-1 $, while the solution at $x\neq 0$ keeps the form of Eqs. (\ref{SF}) or (%
\ref{SDF}), respectively], with total power%
\begin{equation}
P_{\sigma }(k)=2\left[ \frac{1+\gamma ^{2}}{2\varepsilon _{2}}+\sigma \sqrt{%
2k}-\sigma \sqrt{2k+\frac{\left( 1+\gamma ^{2}\right) ^{2}}{4\varepsilon
_{2}^{2}}}\right] .  \label{Power2}
\end{equation}%
Note that expressions (\ref{xi3}) and (\ref{Power2}) depend on the $\mathcal{%
PT}$ strength $\gamma $, unlike their counterparts (\ref{xi2}) (\ref{Power}%
). Furthermore, solution (\ref{xi3}) exists for all values of $k>0$, unlike
the one given by Eq. (\ref{xi2}), whose existence region is limited by
condition (\ref{><}). The numerical analysis demonstrates that these
solutions have a narrow stability area (at small $\gamma $) for the SF bulk
medium, $\sigma =+1$, and are completely unstable for $\sigma =-1$.

\section{Gap solitons supported by a hot spot in the Bragg grating}

As mentioned above, the first example of SDSs supported by an HS in lossy
media was predicted in the framework of the CMEs (\ref{o1}) and (\ref{o2})
for the BG in a nonlinear waveguide \cite{Mak}. It is relevant to outline
this original result in the present review. Unlike the basic models
considered above, the CME system does not admit exact solutions for pinned
solitons, because Eqs. (\ref{o1}) and (\ref{o2}) do not have analytical
solutions in the bulk in the presence of the loss terms, $\gamma >0$,
therefore analytical consideration (verified by numerical solutions) is only
possible in the framework of the perturbation theory, which treats $\gamma $
and $\Gamma _{1}$ as small parameters, while the strength of the local
potential, $\Gamma _{2}$, does not need to be small. This example of the
application of the perturbation theory is important, as it provided a
paradigm for the analysis of other models, where exact solutions are not
available either \cite{Valery,Ye}.

\subsection{The zero-order approximation}

A stationary solution to Eqs. (\ref{o1}) and (\ref{o2}) with $\gamma =\Gamma
_{1}=0$ and $\Gamma _{2}>0$ is sought for in the form which is common for
quiescent BG solitons \cite{Christo}-\cite{Sterke}:
\begin{eqnarray}
u\left( x,z\right) =U(x)\exp \left( -iz\cos \theta \right) ,  \notag \\
v(x,z)=-V^{\ast }(x)\exp \left( -iz\,\cos \theta \right) ,  \label{statasmpt}
\end{eqnarray}%
where $\ast $ stands for the complex conjugate, $\theta $ is a parameter of
the soliton family, and function $U(x)$ satisfies equation%
\begin{equation}
\left[ i\frac{d}{dx}U+\cos \theta +\Gamma _{2}\delta (x)\right]
U+3|U|^{2}U-U^{\ast }=0.  \label{reduced}
\end{equation}%
The integration of Eq. (\ref{reduced}) around $x=0$ yields the respective
b.c., $U(x=+0)=U(x=-0)\exp (i\Gamma _{2})$, cf. Eq. (\ref{bc})$\,$. As shown
in \cite{Mak-earlier}, an exact soliton-like solution to Eq. (\ref{reduced}%
), supplemented by the b.c., can be found, following the pattern of the
exact solution for the ordinary gap solitons \cite{Aceves}-\cite{Sterke}:%
\begin{equation}
U(x)=\frac{1}{\sqrt{3}}\frac{\sin \theta }{\cosh \left[ \left( |x|+\xi
\right) \sin \theta -\frac{i}{2}\theta \right] }\,,  \label{solution}
\end{equation}%
where offset $\xi $, cf. Eqs. (\ref{sinh}), (\ref{SF}), (\ref{SDF}), is
determined by the relation
\begin{equation}
\tanh \left( \xi \sin \theta \right) =\frac{\tan \left( \Gamma _{2}/2\right)
}{\tan \left( \theta /2\right) }\,.  \label{tan}
\end{equation}%
Accordingly, the soliton's squared amplitude (peak power) is
\begin{equation}
\left\vert U(x=0)\right\vert ^{2}=\left( 2/3\right) \left( \cos \Gamma
_{2}-\cos \theta \right) \,.  \label{amplitude}
\end{equation}%
From Eq. (\ref{tan}) it follows that the solution exists not in the whole
interval $0<\theta <\pi $, where the ordinary gap solitons are found, but in
a region determined by constraint $\tanh \left( \xi \sin \theta \right) <1$,
i.e.,
\begin{equation}
\Gamma _{2}<\theta <\pi .  \label{interval}
\end{equation}%
In turn, Eq. (\ref{interval}) implies that the solutions exist only for $%
0\leq \Gamma _{2}<\pi $ ($\Gamma _{2}<0$ corresponds to the repulsive HS,
hence the soliton pinned to it will be unstable).

\subsection{The first-order approximation}

In the case of $\gamma =\Gamma _{1}=0$, Eqs. (\ref{o1}) and (\ref{o2})
conserve the total power,
\begin{equation}
P=\int_{-\infty }^{+\infty }\left[ \left\vert u(x)\right\vert
^{2}+\left\vert v(x)\right\vert \right] ^{2}dx\,.  \label{E}
\end{equation}%
In the presence of the loss and gain, the exact evolution equation for the
total power is
\begin{equation}
\frac{dP}{dz}=-2\gamma P+2\Gamma _{1}\left[ \left\vert u(x)\right\vert
^{2}+\left\vert v(x)\right\vert ^{2}\right] |_{x=0}\,.  \label{dE/dt}
\end{equation}%
If $\gamma $ and $\Gamma _{1}$ are treated as small perturbations, the
balance condition for the power, $dP/dz=0$, should select a particular
solution, from the family of exact solutions (\ref{solution}) of the
conservative model, which remains, to the first approximation, a stationary
pinned soliton, cf. Eq. (\ref{balance}).

The balance condition following from Eq. (\ref{dE/dt}) demands $\gamma
P=\Gamma _{1}\left[ \left\vert U(x=0)\right\vert ^{2}+\left\vert
V(x=0)\right\vert ^{2}\right] \,.$ Substituting, in the first approximation,
the unperturbed solution (\ref{solution}) and (\ref{tan}) into this
condition, it can be cast in the form of
\begin{equation}
\frac{\theta -\Gamma _{2}}{\cos \Gamma _{2}-\cos \theta }=\frac{\Gamma _{1}}{%
\gamma }\,.  \label{complicated}
\end{equation}

\begin{figure}[bph]
\centerline{\scalebox{0.5}{\includegraphics{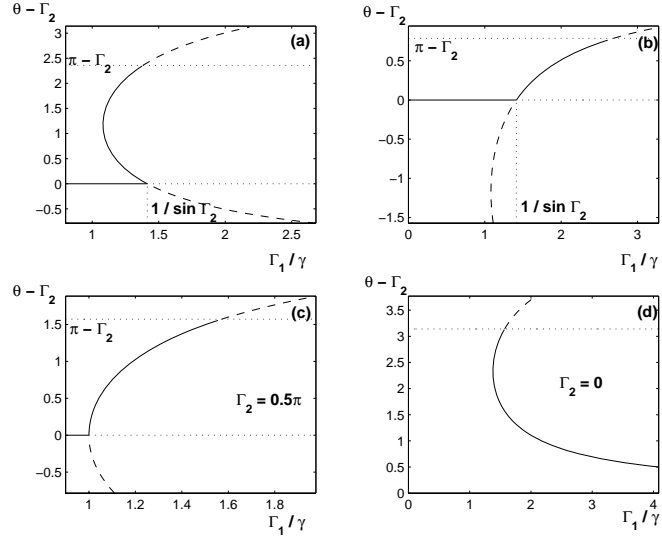}}}
\caption{Analytically predicted solution branches for the pinned gap
soliton
in the BG with weak uniform loss and local gain. Shown is $\protect\theta %
-\Gamma _{2}$ vs. the gain strength, $\Gamma _{1}/\protect\gamma $. (a) $%
\Gamma _{2}=\protect\pi /4$; (b) $\Gamma _{2}=3\protect\pi /4$; (c) $\Gamma
_{2}=\protect\pi /2$; (d) $\Gamma _{2}=0$. In the last case, nontrivial
solutions appear at point $\protect\theta =0.7442\protect\pi ,\,\Gamma _{1}/%
\protect\gamma =1.3801$, and at large values of $\Gamma _{1}/\protect\gamma $
the continuous curve asymptotically approaches the horizontal axis. In all
the panels, the dashed lines show a formal continuation of the solutions in
the unphysical regions, $\protect\theta <\Gamma _{2}$, and $\protect\theta >%
\protect\pi $. In panels (a), (b), and (c), the trivial solution, $\protect%
\theta -\Gamma _{2}=0$, is shown by the solid line where it is stable; in
the case corresponding to the panel (d), all the axis $\protect\theta =0$
corresponds to the stable trivial solution.}
\label{fig9}
\end{figure}

The pinned soliton selected by Eq. (\ref{complicated}) appears, with the
increase of the relative gain strength $\Gamma _{1}/\gamma $, as a result of
a bifurcation. The inspection of Fig. \ref{fig9}, which displays $(\theta
-\Gamma _{2})$ vs. $\Gamma _{1}/\gamma $, as per Eq. (\ref{complicated}),
shows that the situation is qualitatively different for $\Gamma _{2}<\pi /2$
and $\Gamma _{2}>\pi /2$.

In the former case, a \textit{tangent} (saddle-node) bifurcation \cite{bif}
occurs at a minimum value $\left( \Gamma _{1}/\gamma \right) _{\min }$ of
the relative gain, with two solutions existing at $\Gamma _{1}/\gamma >$ $%
\left( \Gamma _{1}/\gamma \right) _{\min }$. Analysis of Eq. (\ref%
{complicated}) demonstrates that, with the variation of $\Gamma _{2}$, the
value $\left( \Gamma _{1}/\gamma \right) _{\min }$ attains an absolute
minimum, $\Gamma _{1}/\gamma =1$, at $\Gamma _{2}=\pi /2$. With the increase
of $\Gamma _{1}/\gamma $, the lower unstable solution branch, that starts at
the bifurcation point [see Fig. \ref{fig9}(a)] hits the limit point $\theta
=\Gamma _{2}$ at $\Gamma _{1}/\gamma =1/\sin \Gamma _{2}$, where it
degenerates into the zero solution, according to Eq. (\ref{amplitude}). The
upper branch generated, as a stable one, by the bifurcation in Fig. \ref%
{fig9}(a) continues until it attains the maximum possible value, $\theta
=\pi $, which happens at
\begin{equation}
\frac{\Gamma _{1}}{\gamma }=\left( \frac{\Gamma _{1}}{\gamma }\right) _{\max
}\equiv \frac{\pi -\Gamma _{2}}{1+\cos \Gamma _{2}}\,.  \label{max}
\end{equation}%
In the course of its evolution, this branch may acquire an instability
unrelated to the bifurcation, see below.

In the case $\Gamma _{2}>\pi /2$, the situation is different, as the
saddle-node bifurcation is imaginable in this case, occurring in the
unphysical region $\theta <\Gamma _{2}$, see Fig. \ref{fig9}(b). The only
physical branch of the solutions appears as a stable one at point $\Gamma
_{1}/\gamma =1/\sin \Gamma _{2}$, where it crosses the zero solution, making
it the unstable. However, as well as the above-mentioned branch, the present
one may be subject to an instability of another type. This branch ceases to
be a physical one at point (\ref{max}). At the boundary between the two
generic cases considered above, i.e., at $\Gamma _{2}=\pi /2$, the
bifurcation occurs exactly at $\theta =\pi /2$, see Fig. \ref{fig9}(c).

The situation is different too in the case $\Gamma _{2}=0$ [see Fig. 1(d)],
when the HS has no local-potential component, and Eq. (\ref{complicated})
takes the form of
\begin{equation}
\frac{\theta }{2\sin ^{2}\left( \theta /2\right) }=\frac{\Gamma _{1}}{\gamma
}\,.
\end{equation}%
In this case, the solution branches do not cross axis $\theta =0$ in Fig. %
\ref{fig9}(d). The lower branch, which asymptotically approaches the $\theta
=0$ axis, is unstable, while the upper one might be stable within the
framework of the present analysis. However, numerical results demonstrate
that, in the case of $\Gamma _{2}=0$, the soliton is always unstable against
displacement from $x=0$.

\subsection{Stability of the zero solution}

As done above for the CGLE model with the HS, see Eqs. (\ref{linearized}) - (%
\ref{lambda}), it is relevant to analyze the stability of the zero
background, and the relation between the onset of the localized instability,
driven by the HS, and the emergence of the stable pinned mode, in the
framework of the CMEs. To this end, a solution to the linearized version of
the CMEs is looked for as

\begin{gather*}
\left\{ u(x,z),v(x,z)\right\} =\left\{ A_{+},B_{+}\right\} e^{\Lambda
z-Kx}\,\;\mathrm{at}\;x>0, \\
\left\{ u(x,z),v(x,z)\right\} =\left\{ A_{-},B_{-}\right\} \,e^{\Lambda
z+Kx},\;\mathrm{at}\;x<0,
\end{gather*}%
with \textrm{Re}$\{K\}>0$. A straightforward analysis makes it possible to
eliminate constant $K$ and find the instability growth rate \cite{Mak}:
\begin{equation}
\mathrm{Re}\,\Lambda =-\gamma +\left( \sinh \,\Gamma _{1}\right) \,|\sin
\,\Gamma _{2}|\,.  \label{linrel3}
\end{equation}%
Thus, the zero solution is subject to the HS-induced instability, provided
that the local gain is strong enough:
\begin{equation}
\sinh \,\Gamma _{1}>\sinh \left( \left( \Gamma _{1}\right) _{\mathrm{cr}%
}\right) \equiv \gamma /|\sin \,\Gamma _{2}|\,.  \label{critical}
\end{equation}%
Note that the instability is impossible in the absence of the local
potential $\sim \Gamma _{2}$. The instability-onset condition (\ref{critical}%
) simplifies in the limit when both the loss and gain parameters are small
(while $\Gamma _{2}$ is not necessarily small): $\,$%
\begin{equation}
\Gamma _{1}>\left( \Gamma _{1}\right) _{\mathrm{cr}}\approx \gamma /|\sin
\,\Gamma _{2}|\,\,.  \label{cr2}
\end{equation}%
As seen in Fig. \ref{fig9}, the pinned mode emerges at critical point (\ref%
{cr2}), for $\Gamma _{2}\geq \pi /2$, while at $\Gamma _{2}<\pi /2$ it
emerges at $\Gamma _{1}<$ $\left( \Gamma _{1}\right) _{\mathrm{cr}}$.

\subsection{Numerical results}

First, direct simulations of the conservative version of CMEs (\ref{o1}), (%
\ref{o2}), with $\gamma =\Gamma _{1}=0$ (and $\Gamma _{2}>0$) and an
appropriate approximation for the delta-function, have demonstrated that
there is a very narrow stability interval in terms of parameter $\theta $
[see Eqs. (\ref{reduced})-(\ref{interval}), close to (slightly larger than) $%
\theta _{\mathrm{stab}}\approx \pi /2$, such that the solitons with $\theta
<\theta _{\mathrm{\ stab}}$ decay into radiation, while initially created
solitons with $\theta >\theta _{\mathrm{stab}}$ spontaneously evolve,
through emission of radiation waves, into a stable one with $\theta \approx
\theta _{\mathrm{stab}}$ \cite{Mak-earlier,Mak}. This value weakly depends
on $\Gamma _{2}$. For instance, at $\Gamma _{2}=0.1$, the stability interval
is limited to $0.49\pi <\theta <0.52\pi $, while at a much larger value of
the local-potential strength, $\Gamma _{2}=1.1$, the interval is located
between boundaries $0.51\pi <\theta <0.55\pi $. In this connection, it is
relevant to mention that in the usual BG model, based on Eqs. (\ref{o1}), (%
\ref{o2}) with $\Gamma _{2}=0$, the quiescent solitons are stable at $\theta
<\theta _{\mathrm{cr}}^{(0)}\approx 1.011\cdot \left( \pi /2\right) $ \cite%
{Barash,Trillo}.

Direct simulations of the full CME system (\ref{o1}), (\ref{o2}), which
includes weak loss and local gain, which may be considered as perturbations,
produce stable dissipative gap solitons with small but persistent internal
oscillations, i.e., these are, strictly speaking, breathers, rather than
stationary solitons \cite{Mak}. An essential finding is that the average
value of $\theta $ in such robust states may be essentially larger than the
above-mentioned $\theta _{\mathrm{stab}}\approx \pi /2$ selected by the
conservative counterpart of the system, see particular examples in Table 1.
With the increase of the gain strength $\Gamma _{1}$, the amplitude of the
residual oscillations increases, and, eventually, the oscillatory state
becomes chaotic, permanently emitting radiation waves, which increases the
effective loss rate that must be compensated by the local gain.

Taking values of $\gamma $ and $\Gamma _{1}$ small enough, it was checked if
the numerically found solutions are close to those predicted by the
perturbation theory in the form of Eqs. (\ref{solution}) and (\ref{tan}),
with $\theta $ related to $\gamma $ and $\Gamma _{1}$ as per Eq. (\ref%
{complicated}). It has been found that the quasi-stationary solitons (with
the above-mentioned small-amplitude intrinsic oscillations) are indeed close
to the analytical prediction. The corresponding values of $\theta $ were
identified by means of the best fit to expression~(\ref{solution}). Then,
for the so found values of $\theta $ and given loss coefficient $\gamma $,
the equilibrium gain strength $\Gamma _{1}$ was calculated as predicted by
the analytical formula (\ref{complicated}), see a summary of the results in
Table 1. It is seen in the Table that the numerically found equilibrium
values of $\Gamma _{1}$ exceeds the analytically predicted counterparts by $%
\sim 10\%-15\%$, which may be explained by the additional gain which is
necessary to compensate the permanent power loss due to the emission of
radiation.

\begin{tabular}[t]{|l|l|l|l|l|}
\hline
$\gamma $ & $\theta $ & $\left( \Gamma _{1}\right) _{\mathrm{num}}$ & $%
\left( \Gamma _{1}\right) _{\mathrm{anal}}$ & $\frac{\left( \Gamma
_{1}\right) _{\mathrm{num}}-\left( \Gamma _{1}\right) _{\mathrm{anal}}}{%
\left( \Gamma _{1}\right) _{\mathrm{anal}}}$ \\ \hline
$0.000316$ & $0.5\pi $ & $0.000422$ & $0.000386$ & $0.0944$ \\ \hline
$0.00316$ & $0.595\pi $ & $0.0042$ & $0.00369$ & $0.1373$ \\ \hline
$0.01$ & $0.608\pi $ & $0.01333$ & $0.01165$ & $0.1442$ \\ \hline
$0.1$ & $0.826\pi $ & $0.1327$ & $0.121$ & $0.0967$ \\ \hline
\end{tabular}

{\small Table 1. Values of the loss parameter $\gamma $ at which
quasi-stationary stable pinned solitons were found in simulations of the CME
system, (\ref{o1}), (\ref{o2}), by adjusting gain $\Gamma _{1}$, for fixed $%
\Gamma _{2}=0.5$. Values of the soliton parameter, $\theta $, which provide
for the best fit of the quasi-stationary solitons to the analytical solution
(\ref{solution}) are displayed too. $\left( \Gamma _{1}\right) _{\mathrm{anal%
}}$ is the gain coefficient predicted, for given $\gamma $, $\theta $, and $%
\Gamma _{2}$, by the energy-balance equation (\ref{complicated}), which does
not take the radiation loss into account. The rightmost column shows the
relative difference between the numerically found and analytically predicted
gain strength, which is explained by the necessity to compensate additional
radiation loss.}

It has also been checked that the presence of nonzero attractive potential
with $\Gamma _{2}>0$ is necessary for the stability of the gap solitons
pinned to the HS in the BG model. In addition to the analysis of stationary
pinned modes, collisions between\ a gap soliton, freely moving in the weakly
lossy medium, with the HS were also studied by means of direct simulations
\cite{Mak}. The collision splits the incident soliton into the transmitted
and trapped components.

\section{Discrete solitons pinned to the hot spot in the lossy lattice}

The 1D version of the discrete model based on Eq. (\ref{eq:gl}), i.e.,
\begin{eqnarray}
\frac{du_{n}}{dz}=\frac{i}{2}\left( u_{n-1}+u_{n+1}+u_{n-1}-2u_{m,n}\right)
\notag \\
-\gamma u_{n}+\left[ \left( \Gamma _{1}+i\Gamma _{2}\right) +\left(
iB-E\right) |u_{n}|^{2}\right] \delta _{n,0}u_{n},  \label{1D}
\end{eqnarray}%
makes it possible to gain insight into the structure of lattice solitons
supported by the ``hot site", at which both the gain and nonlinearity are
applied, as in that case an analytical solution is available \cite{we}. It
is relevant to stress that the present model admits stable localized states
even in the case of the unsaturated cubic gain, $E<0$, see details below.

\subsection{Analytical results}

The known \textit{staggering transformation} \cite{review,PGK}, $%
u_{m}(t)\equiv \left( -1\right) ^{m}e^{-2it}\tilde{u}_{m}^{\ast }$, where
the asterisk stands for the complex conjugate, reverses the signs of $\Gamma
_{2}$ and $B$ in Eq. (\ref{1D}). Using this option, the signs are fixed by
setting $\Gamma _{2}>0$ (the linear discrete potential is attractive), while
$B=+1$ or $B=-1$ corresponds to the SF and SDF nonlinearity, respectively.
Separately considered is the case of $B=0$, when the nonlinearity is
represented solely by the cubic dissipation localized at the HS.

Dissipative solitons with real propagation constant $k$ are looked for by
the substitution of $u_{m}=U_{m}e^{ikz}\ $in Eq. (\ref{1D}). Outside of the
HS site, $m=0$, the linear lattice gives rise to the exact solution with
real amplitude $A$,
\begin{equation}
U_{m}=A\exp (-\lambda |m|),\;|m|~\geq 1,  \label{eq:exp}
\end{equation}%
and complex $\lambda \equiv \lambda _{1}+i\lambda _{2}$, localized modes
corresponding to $\lambda _{1}>0$. The analysis demonstrates that the
amplitude at the HS coincides with $A$, i.e., $U_{0}=A$. Then, the remaining
equations at $n=0$ and $n=1$ yield a system of four equations for four
unknowns, $A$, $\lambda _{1}$, $\lambda _{2}$, and $k$:
\begin{gather}
-1+\cosh \lambda _{1}\cos \lambda _{2}=k,~-\gamma -\sinh \lambda _{1}\sin
\lambda _{2}=0,  \notag \\
e^{-\lambda _{1}}\sin \lambda _{2}-\gamma +\Gamma _{1}-EA^{2}=0,  \notag \\
e^{-\lambda _{1}}\cos \lambda _{2}-1+\Gamma _{2}+BA^{2}=k.  \label{eq:sys}
\end{gather}%
This system was solved numerically. The stability of the discrete SDSs was
analyzed by the computation of eigenvalues for modes of small perturbations
\cite{PGK,Yang} and verified by means of simulations of the perturbed
evolution. Examples of stable discrete solitons can be seen below in Fig. %
\ref{fig12}.

In the linear version of the model, with $B=E=0$ in Eq. (\ref{1D}),
amplitude $A$ is arbitrary, dropping out from Eqs. (\ref{eq:sys}). In this
case, $\Gamma _{1}$ may be considered as another unknown, determined by the
balance between the background loss and localized gain, which implies
structural instability of the stationary trapped modes in the linear model
against small variations of $\Gamma _{1}$. In the presence of the
nonlinearity, the power balance is adjusted through the value of the
amplitude at given $\Gamma _{1}$, therefore solutions can be found in a
range of values of $\Gamma _{1}$. Thus, families of pinned modes can be
studied, using linear gain $\Gamma _{1}$ and cubic gain/loss $E$ as control
parameters (in the underlying photonic lattice, their values may be adjusted
by varying the intensity of the external pump).

\subsection{Numerical results}

\subsubsection{The self-defocusing regime ($B=-1$)}

\begin{figure}[t]
\begin{center}
\includegraphics[width = 58mm, keepaspectratio]{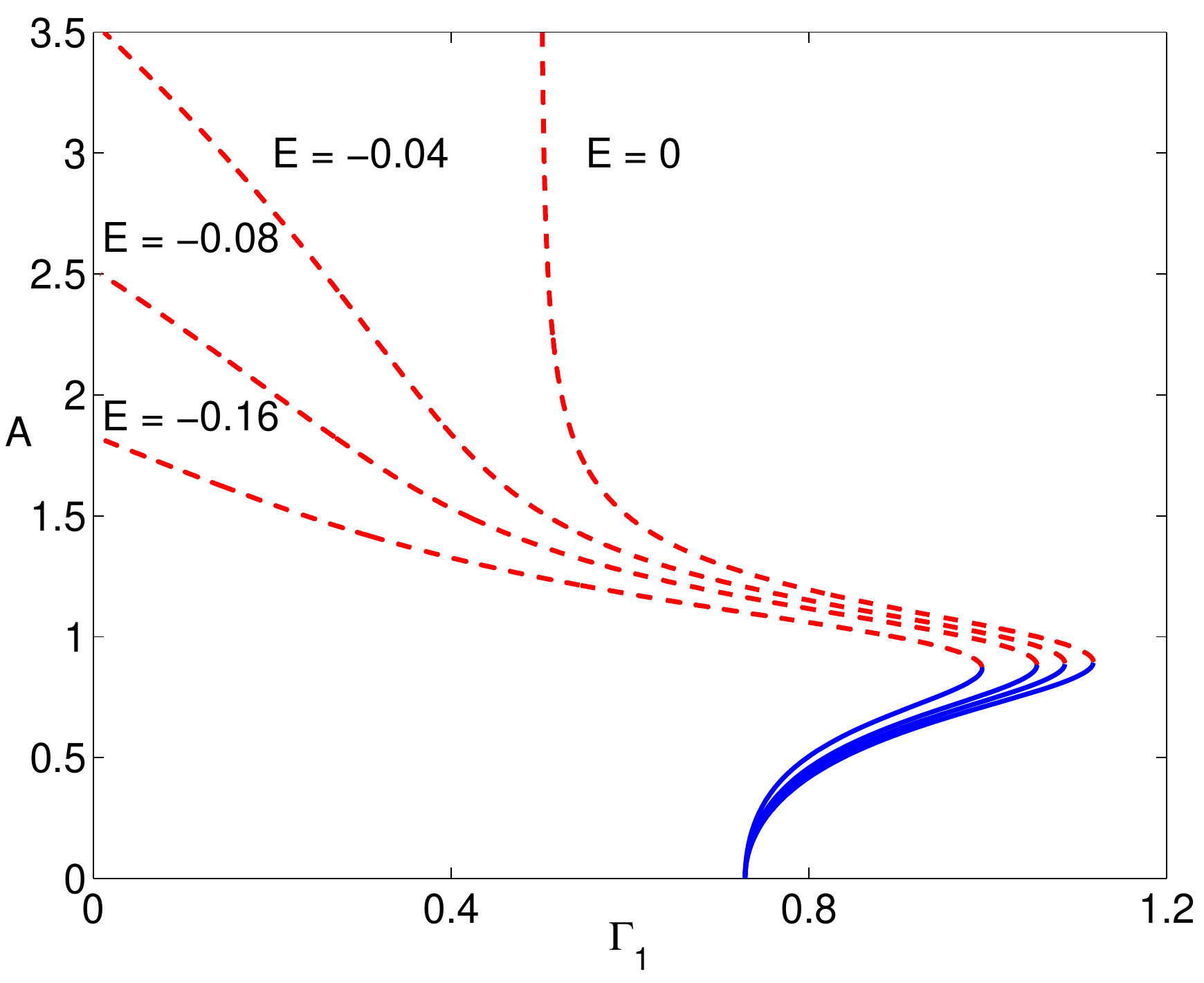}
\end{center}
\caption{(Color online) Amplitude $A$ of the pinned 1D discrete soliton as a
function of linear ($\Gamma _{1}$) and cubic ($E\leq 0$) gain. Other
parameters in Eq. (\protect\ref{1D}) are $\protect\gamma =0.5$, $\Gamma
_{2}=0.8$, and $B=-1$ (the SDF nonlinearity). Here and in Figs. \protect\ref%
{fig11} and \protect\ref{fig13} below, continuous and dashed curves denote
stable and unstable solutions, respectively.}
\label{fig10}
\end{figure}

The most interesting results were obtained for the \emph{unsaturated
nonlinear gain}, i.e., $E<0$ in Eq. (\ref{1D}). Figure~\ref{fig10} shows
amplitude $A$ of the stable (solid) and unstable (dashed) pinned modes as a
function of linear gain $\Gamma _{1}>0$ and cubic gain. The existence of
stable subfamilies in this case is a noteworthy finding. In particular, at $%
E=0$ there exists a family of stable pinned modes in the region of $0.73\leq
\Gamma _{1}\leq 1.11$, with the amplitude ranging from $A=0.08$ to $A=0.89$.
Outside this region, solutions decay to zero if the linear gain is too weak (%
$\Gamma _{1}<0.73$), or blow up at $\Gamma _{1}>1.11$. Figure~\ref{fig10}
shows that the bifurcation of the zero solution $A=0$ into the pinned mode
takes place at a particular value $\Gamma _{1}=0.7286$, which is selected by
the above-mentioned power-balance condition in the linear system.

\begin{figure}[t]
\begin{center}
\includegraphics[width = 58mm, keepaspectratio]{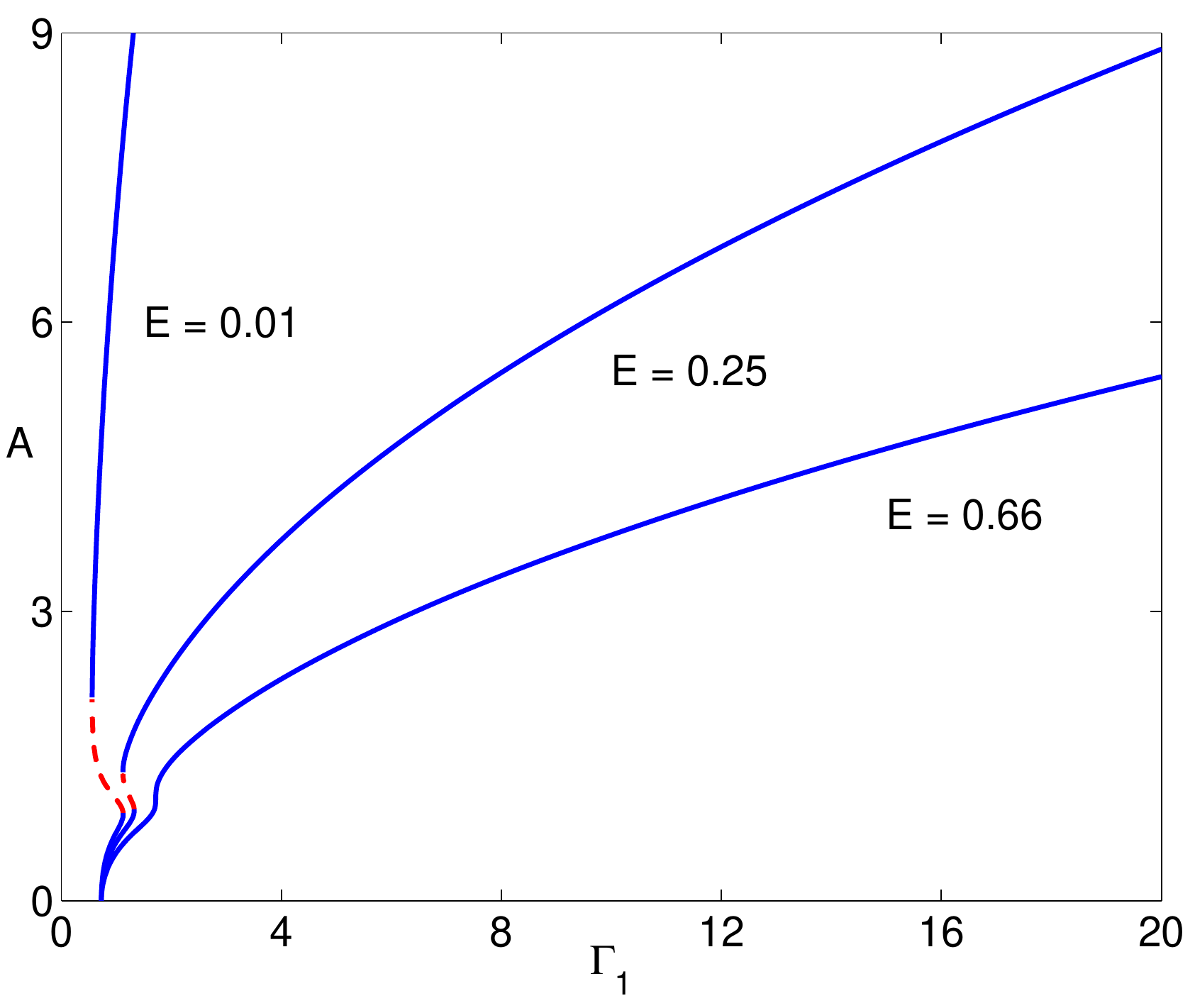}
\end{center}
\caption{(Color online) Solution branches for the discrete solitons in the
case of the cubic loss ($E>0$) and SDF nonlinearity ($B=-1$). The other
parameters are same as those in Fig.~\protect\ref{fig10}.}
\label{fig11}
\end{figure}

\begin{figure}[t]
\begin{center}
\includegraphics[width = 70mm, keepaspectratio]{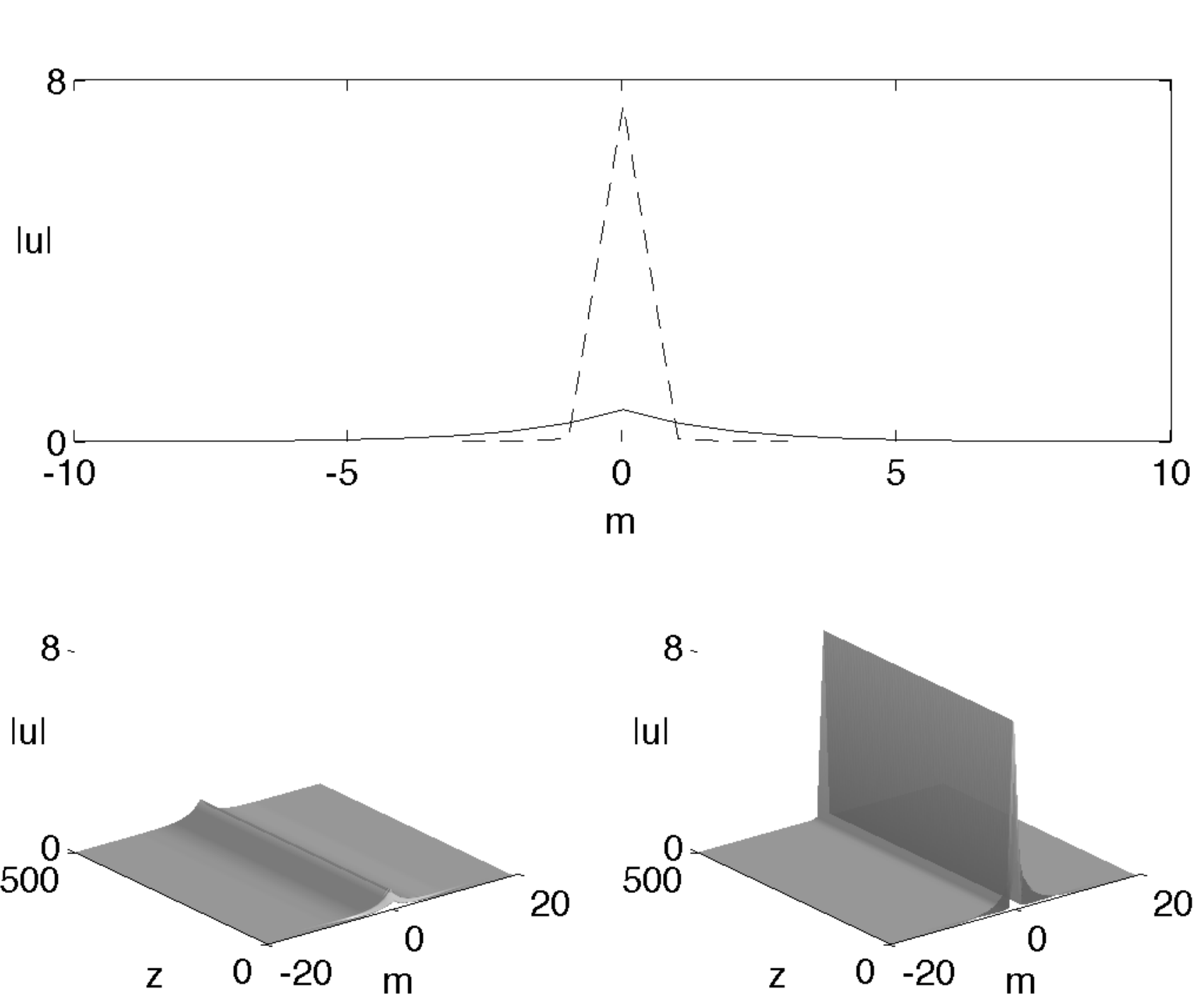}
\end{center}
\caption{The coexistence of stable small- and large-amplitude pinned
discrete modes (top), and the corresponding evolution of nonstationary
solutions (bottom) at $E=0.01$, in the bistability region. Inputs with
amplitudes $A=0.3$ and $A=2$ evolve into the small-amplitude and
large-amplitude stationary modes, respectively. The other parameters are $%
\protect\gamma =0.5$, $\Gamma _{1}=1$, $\Gamma _{2}=0.8$, and $B=-1$ (the
SDF nonlinearity).}
\label{fig12}
\end{figure}

The existence of the stable pinned modes in the absence of the gain
saturation being a remarkable feature, the stability region is, naturally,
much broader in the case of the cubic loss, $E>0$. Figure~\ref{fig11} shows
the respective solution branches obtained with the SDF nonlinearity. When
the cubic loss is small, e.g., $E=0.01$, there are two distinct families of
stable modes, representing broad small-amplitude ($A\leq 0.89$) and narrow
large-amplitude ($A\geq 2.11$) ones. These two stable families are linked by
an unstable branch with the amplitudes in the interval of $0.89<A<2.11$. The
two stable branches coexist in the interval of values of the linear gain $%
0.73\leq \Gamma _{1}\leq 1.13$, where the system is \emph{bistable}. Figure~%
\ref{fig12} shows an example of the coexisting stable modes in the
bistability region. In simulations, a localized input evolves into either of
these two stable solutions, depending on the initial amplitude.

While the pinned modes may be stable against small perturbations under the
combined action of the self-defocusing nonlinearity ($B=-1$) and unsaturated
nonlinear gain ($E\leq 0$), one may expect fragility of these states against
finite-amplitude perturbations. It was found indeed that sufficiently strong
perturbations destroy those modes \cite{we}.

\subsubsection{The self-focusing regime ($B=+1$)}

\begin{figure}[t]
\begin{center}
\includegraphics[width = 58mm, keepaspectratio]{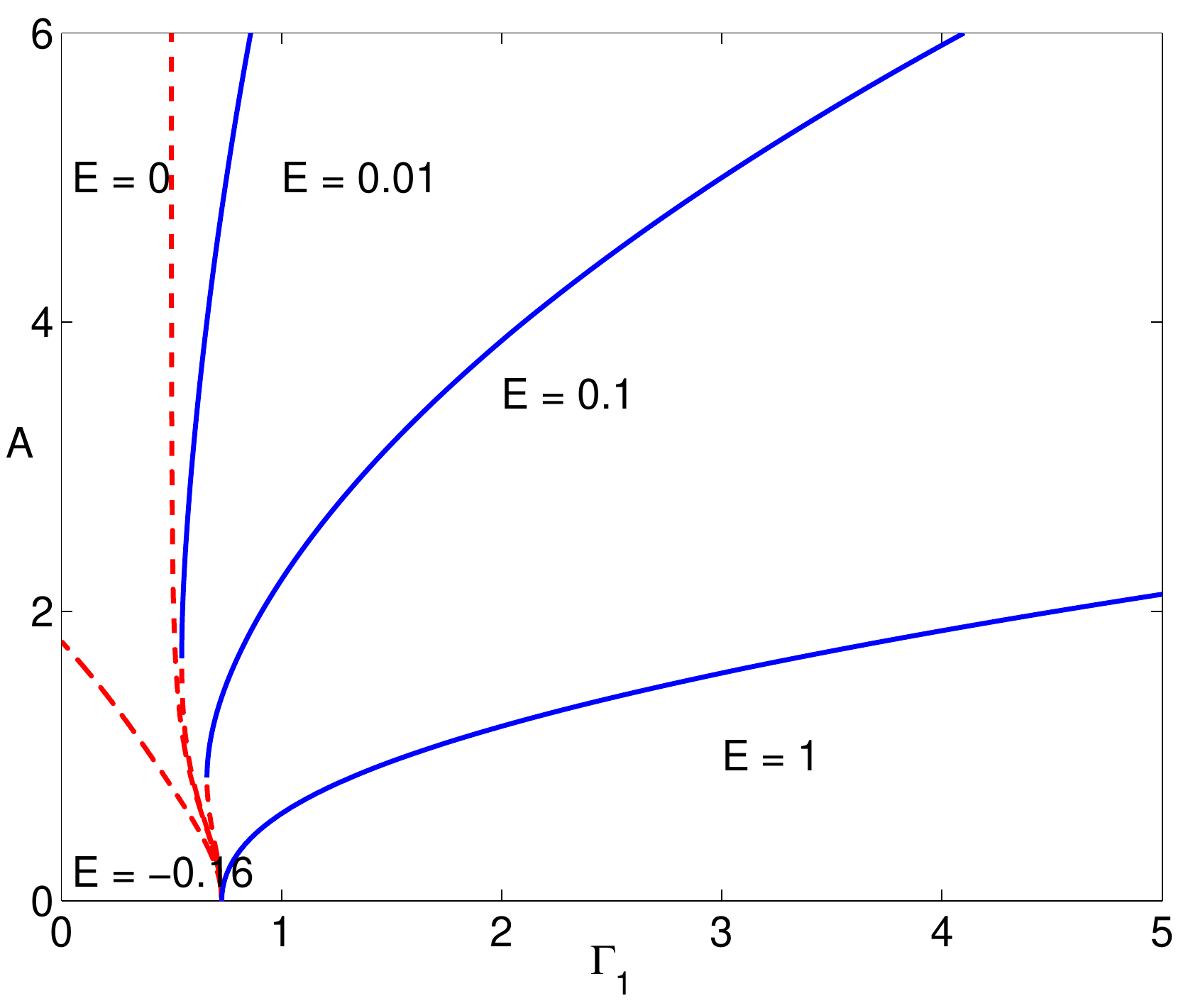}
\end{center}
\caption{(Color online) Amplitude $A$ of the discrete pinned mode as a
function of the linear gain ($\Gamma _{1}$) and cubic loss ($E$) in the case
of the SF nonlinearity, $B=+1$. The other parameters are $\protect\gamma %
=0.5 $ and $\Gamma _{2}=0.8$.}
\label{fig13}
\end{figure}

In the case of the SF nonlinearity, $B=+1$, the pinned-mode branches~are
shown in Fig.~\ref{fig13}, all of them being unstable without the cubic
loss, i.e., at $E\leq 0$. For the present parameters, all the solutions
originate, in the linear limit, from the power-balance point $\Gamma
_{1}=0.73$. The localized modes remain stable even at very large values of $%
\Gamma _{1}$. Lastly, in the case of $B=0$, when the nonlinearity is
represented solely by the cubic gain or loss, all the localized states are
unstable under the cubic gain, $E<0$, and stable under the cubic loss, $E>0$.

\section{Conclusion}

This article presents a limited review of theoretical results obtained in
recently studied 1D and 2D models, which predict a generic method for
supporting stable spatial solitons in dissipative optical media, based on
the use of the linear gain applied in narrow active segments (HSs, ``hot
spots") implanted into the lossy waveguide. In some cases, the unsaturated
cubic localized gain may also support stable spatial solitons, which is a
counter-intuitive finding. In view of the limited length of the article, it
combined a review of the broad class of such models with a more detailed
consideration of selected 1D models where exact or approximate analytical
solutions for the dissipative solitons are available. Naturally, the
analytical solutions, in the combination with their numerical counterparts,
provide a deeper understanding of the underlying physical models.

A relevant issue for the further development of the topic is a possibility
of the existence of \emph{asymmetric modes} supported by symmetric double
HSs, i.e., the analysis of the spontaneous symmetry breaking in this setting
\cite{book}. Thus far, such a result was not demonstrated in a clear form.
Another challenging extension may be a possibility of \emph{dragging} a
pinned 1D or 2D soliton by a HS \emph{moving} across a lossy medium.

\section{Acknowledgement}

I appreciate the invitation from Prof. G. Swartzlander, the Editor-in-Chief
of J. Opt. Soc. Am. B, to draft and submit this mini-review. My work on the
present topic has greatly benefited from collaborations with many
colleagues, including P. L. Chu (deceased), N. B. Aleksi\'{c}, J. Atai, O.
V. Borovkova, K. W. Chow, P. Colet, M. C. Cross, E. Ding, R. Driben, W. J.
Firth, D. Gomila, Y.-J. He, Y. V. Kartashov, E. Kenig, S. K. Lai, C. K. Lam,
H. Leblond, Y. Li, R. Lifshitz, V. E. Lobanov, W. C. K. Mak, A. Marini, T.
Mayteevarunyoo, D. Mihalache, A. A. Nepomnyashchy, P. V. Paulau, H.
Sakaguchi, V. Skarka, D. V. Skryabin, A.\ Y. S. Tang, L. Torner, C. H.
Tsang, V. A. Vysloukh, P. K. A. Wai, H. G. Winful, F. Ye, and X. Zhang.

\end{document}